\documentclass[12pt]{article}

\usepackage[numbers, sort&compress]{natbib}
\usepackage{fullpage}
\usepackage{epsfig}
\usepackage{latexsym,amsmath,amssymb,amsthm}
\usepackage{subfig}
\newcommand{\ceil}[1]{\left\lceil {#1} \right\rceil}

\def\maps11{\stackrel {1-1}{\longmapsto}}

\renewcommand{\qed}{\hfill $\blacksquare$}

\begin{document}


\title{Optimizing TCP Goodput and Delay in next generation IEEE 802.11 (ax) devices}

\author{%
Oran Sharon
\thanks{Corresponding author: oran@netanya.ac.il, Tel: 972-4-9831406,
Fax: 972-4-9930525} \\
Department of Computer Science \\
Netanya Academic College \\
1 University St. \\
Netanya, 42365 Israel
\and
Yaron Alpert\\
Intel\\
13 Zarchin St.\\
Ra'anana, 43662, Israel\\
Yaron.alpert@intel.com
}


\date{}

\maketitle

\begin{abstract} 
In this paper we suggest three scheduling strategies for
the IEEE 802.11ax transmission of 
DL unidirectional TCP data
from the Access Point to stations. Two
strategies are based on the Single User operation mode
and one is based on the Multi User operation mode, using
Multi User Multiple-Input-Multiple-Output (MU-MIMO) and
OFDMA. We measure the Goodput of the system as a function
of the time intervals over which these Goodputs are
received in all three strategies. For up to 8
stations the MU strategy outperforms the SU. For 16
and 32 stations it is not clear whether MU outperforms SU
or vice versa. For 64 stations the SU strategies outperform
the MU significantly. 
We also checked the influence of the {\it Delayed Acks} feature
on the received Goodputs and found that this feature has
significance only when the TCP data segments are relatively
short.
\end{abstract}

\bigskip

\noindent
\textbf{Keywords}: 802.11ax; TCP; Aggregation; Reverse Direction; Transmission Opportunity; Goodput; MIMO; Multi User; OFDMA;

\renewcommand{\baselinestretch}{1.3}
\small\normalsize


\section{Introduction}

\subsection{Background}

The latest IEEE 802.11 Standard (WiFi)~\cite{IEEEBase1}, 
created and maintained by 
the IEEE LAN/MAN Standards Committee (IEEE 802.11),
is currently the most effective solution within the range of Wireless Local
Area Networks (WLAN). Since its first release 
in 1997 the standard provides the basis 
for Wireless network products using 
the WiFi brand, and has since been 
improved upon in many ways. 
One of the main goals of these improvements
is to increase the system throughput 
provided by users and to improve
the standard's Quality-of-Service (QoS) capabilities. 
To fulfill the promise of increasing 
IEEE 802.11 performance and QoS capabilities,
a new amendment (IEEE 802.11ax - also
known as High Efficiency (HE) ) was recently 
introduced~\cite{IEEEax}. IEEE 802.11ax is 
considered to be the sixth generation 
of a WLAN in the IEEE 802.11 set 
of WLAN types and is 
a successor to IEEE 802.11ac~\cite{IEEEac,PS}.
The scope of the IEEE 802.11ax amendment is to
define modifications for both the IEEE 802.11 PHY and MAC
layers that enable at least four-fold improvement
in the average throughput per station in densely
deployed networks~\cite{KKL, AVA, DCC, B}.
Currently IEEE 802.11ax project is 
finalizing revision 2.0, which will be the
baseline for WFA IEEE 802.11ax certification.

\subsection{Research question}

In order to achieve its goals, one of 
the main challenges of IEEE 802.11ax is to enable 
UL and DL simultaneous
transmissions by several stations and to improve
Quality-of-Service performance.
The current paper is a 
continuation to papers~\cite{SA1,SA2,SA3}.
In these papers the authors suggest
scheduling strategies for the parallel transmissions of the AP to
a given set of stations using new features
of IEEE 802.11ax . The authors assume UDP-like
traffic where the AP transmits data MSDUs to the stations,
which reply with MAC acknowledgments. 
In this paper  we
assume a DL unidirectional TCP-like traffic in which the AP transmits
TCP Data MSDUs to a given set of stations, and the stations reply
with TCP Ack MSDUs. As far as we know the issue of transmitting
TCP traffic over IEEE 802.11ax has not yet been investigated. 
We suggest several scheduling
strategies for the transmissions of TCP traffic over
the DL using
Single User (SU) and Multi User (MU) modes
for 1, 4, 8, 16, 32 and 64 
stations scenarios over a reliable
channel. 
This is one of the aspects to compare between new
amendments of the IEEE 802.11 standard~\cite{KCC}.
In this paper we are interested in finding an
upper bound on the maximum 
DL unidirectional TCP Goodput
that can be achieved by IEEE 802.11ax and 
comparing between the various scheduling strategies.
Therefore, we assume
the traffic saturation model
where TCP connections always have data to transmit
and the TCP Ack is generated immediately by receivers.
Second, we neutralize any aspects
of the PHY layer as the
number of Spatial Streams (SS) in use and channel correlation
when using 
Multi User Multiple Input Multiple Output 
(MU-MIMO), the use in the sounding protocol etc.

\indent
As mentioned, we assume that every TCP connection has an unlimited number
of TCP Data segments to transmit, and we assume that
transmissions are made using an optimized
(in terms of overhead reduction) two level aggregation scheme
to be described later.
Our goal
is to find an upper bound on
the maximum possible Goodput that the
wireless channel enables the TCP connections,
where the TCP itself does not impose any limitations
on the offered load, i.e. on the rate that MSDUs are
given for transmission to the
MAC layer of the IEEE 802.11ax.
We also assume that the AP and the stations
are the end points of the TCP connections.
Following e.g.~\cite{MKA,BCG1,BCG2,KAMG}
it is quite common to consider short Round Trip Times (RTT)
in this kind of high speed network such that
no retransmission timeouts occur. 
Moreover, we assume that every TCP connections' Transmission Window
can always provide as many MSDUs to transmit as the IEEE 802.11ax
protocol limits
enable. This assumption follows the observation
that aggregation is useful in a scenario where the offered
load on the channel is high. 
Finally, we assume that every TCP Ack either acknowledges 
one TCP Data segment, or it 
acknowledges two TCP Data segments. The latter possibility
is denoted {\it Delayed Acks},
a feature in TCP that enables a TCP Ack to acknowledge two TCP
Data segments.

This research is only a first step in investigating TCP traffic
in IEEE 802.11ax. In our further papers we plan
to address other 
TCP traffic scenarios to investigate
such
as UL unidirectional TCP traffic and bi-directional TCP traffic.

\subsection{Previous works}

The issue of TCP traffic over IEEE 802.11ax 
that involves bidirectional data packet exchange
has not yet been studied.
Most of the research papers on IEEE 802.11ax
thus far examine different
access methods to enable efficient multi-user access
to random sets of stations.
For example, 
in~\cite{QLYY} the authors deal with the introduction
of Orthogonal Frequency Division Multiple Access (OFDMA)
into IEEE 802.11ax to enable multi user access.
They introduce an OFDMA based multiple access
protocol, denoted Orthogonal MAC for IEEE 802.11ax (OMAX),
to solve synchronization
problems and reduce overhead associated with
using OFDMA.
In~\cite{LLYQYZY} the authors suggest an access protocol
over the UL of an IEEE 802.11ax WLAN based on
MU-MIMO and OFDMA PHY.
In~\cite{LDC} the authors suggest a centralized medium
access protocol for the UL of IEEE 802.11ax in order to efficiently
use the transmission resources.
In this protocol, stations transmit
requests for frequency sub-carriers, denoted
Resource Units (RU), to the AP over the UL. The AP
allocates RUs to the stations which later use them
for data transmissions over the UL.
In~\cite{KBPSL} a new method to use OFDMA over the UL
is suggested, where MAC Protocol Data Units 
(MPDU) from the stations are
of different lengths.
In~\cite{JS, RFBBO, RBFB, HYSG} a new version 
of the Carrier Sense Multiple Access with Collision Avoidance
(CSMA/CA) protocol, denoted Enhanced CSMA/CA (CSMA/ECA) is
suggested for MU transmissions, 
which is suitable for IEEE 802.11ax . A deterministic
BackOff is used after a successful transmission, and the BackOff
stage is not reset after service. The BackOff stage is reset
only when a station does not have any further MPDUs to transmit.
CSMA/ECA enables more efficient use of the channel
and enhanced fairness.
In~\cite{KLL} the authors assume
a network with legacy and IEEE 802.11ax stations and examine
fairness issues between the two sets of stations. 

\indent
We would like to mention that the issue of TCP traffic
over IEEE 802.11ac networks (the predecessor standard of IEEE 802.11ax)
has already been investigated, e.g.
in~\cite{SA10,SA11,SA12}, for DL TCP traffic, UL TCP traffic
and both DL and UL TCP traffic.
However, in all these works there is no possibility
of using the MU operation mode over the UL, a feature
that was first introduced in IEEE 802.11ax .

\indent
The remainder of the paper is organized as follows: 
In Section 2 we describe
the new mechanisms of IEEE 802.11ax 
relevant to this paper. In Section 3 we 
describe the scheduling strategies that
we suggest
in SU and MU modes.
We assume the reader is familiar 
with the basics of PHY and MAC layers
of IEEE 802.11 described in previous papers, e.g.~\cite{SA}. 
In Section 4 we analytically compute 
the Goodputs of the various scheduling strategies.
In Section 5 we present the Goodputs
of the various scheduling strategies and
Section 6 summarizes 
the paper. In the Appendix we show how
to efficiently schedule MPDUs in the various
scheduling strategies.
Lastly, moving forward,
we denote IEEE 802.11ax 
by 11ax .

\section{The new features in IEEE 802.11ax}

IEEE 802.11ax focuses on implementing 
mechanisms to efficiently serve more
users, enabling consistent and 
reliable streams of data ( average throughput
per user ) in the presence of multiple users. 
In order to meet these targets 11ax addresses several 
new mechanisms in both the PHY and
MAC layers. At the PHY layer, 
11ax enables larger OFDM FFT sizes (4X larger)
and therefore every OFDM symbol is 
$12.8 \mu s$ compared to $3.2 \mu s$ in 
IEEE 802.11ac, the predecessor of 11ax .
By narrower sub-carrier spacing (4X closer)
the protocol efficiency is increased because
the same Guard Interval (GI) is used both in 11ax and 
in previous versions of the standard.

In addition, to increase the average 
throughput per user in high-density scenarios,
11ax introduces two new Modulation Coding Schemes 
(MCSs), MCS10 (1024 QAM ) and MCS 11 (1024 QAM 5/6), 
applicable for transmission with bandwidth larger than 20 MHz.

In this paper we use the Transmission Opportunity (TXOP)
feature first introduced in IEEE 802.11n~\cite{IEEEn}.
This feature allows a station, after gaining
access to the channel, to transmit
several PHY Protocol Data Units
(PPDUs) in a row without interruption, and can also allocate
some of the TXOP time interval to one or more receivers
in order to allow data transmission in the reverse link.
This is termed {\it Reverse Direction (RD)}. For scenarios
with bidirectional traffic such as TCP Data segments/Ack segments,
this approach is very efficient as it reduces contention
in the wireless channel.

We focus on optimizing 
the TXOP duration and pattern, PPDU duration and
the 11ax's two-level aggregation
scheme working point first introduced 
in IEEE 802.11n~\cite{IEEEn}, 
in which several  MPDUs can be aggregated 
to be transmitted in a  single PHY Service 
Data Unit (PSDU). Such aggregated PSDU 
is denoted Aggregate MAC Protocol Data
Unit (A-MPDU) frame. In two-level aggregation 
every MPDU can contain several 
MAC Service Data Units (MSDU). 
MPDUs are separated by an MPDU 
Delimiter field of 4 bytes and 
each MPDU contains MAC Header
and Frame Control Sequence (FCS) fields.
MSDUs within an MPDU are separated 
by a SubHeader field of 14 bytes. 
Every MSDU is rounded to an 
integral multiple of 4 bytes 
together with the SubHeader field. 
Every MPDU is also rounded to 
an integral multiple of 4 bytes.

\begin{figure}
\vskip 9cm
\includegraphics{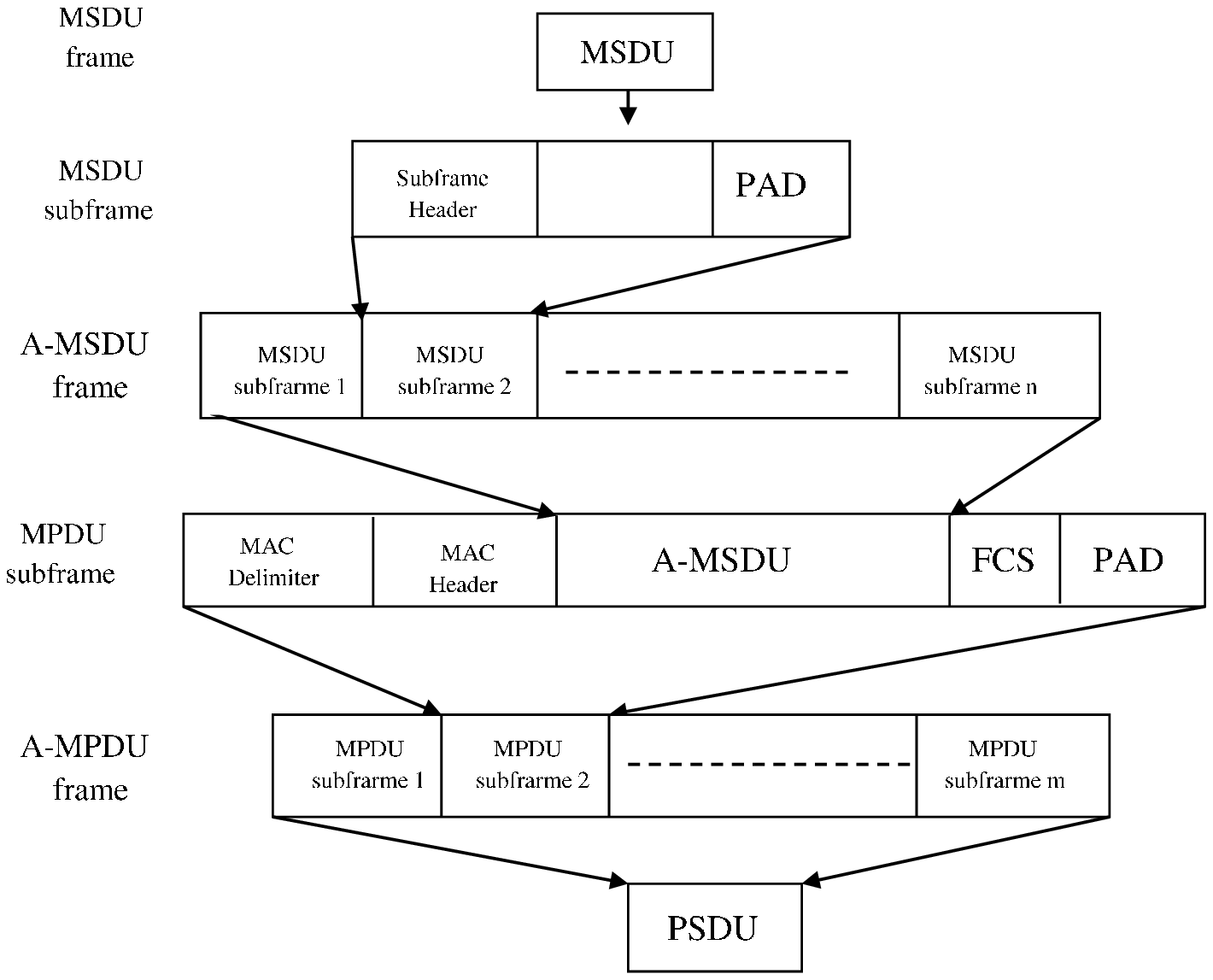}
\caption{The generation of an A-MPDU frame
in two-level aggregation.}
\label{fig:twole}
\end{figure}

In 11ax the size of 
an MPDU is limited to 11454 bytes and the
size of the A-MPDU frame is 
limited to 4,194,304 bytes.
The transmission time of the PPDU 
(PSDU and its preamble) is limited to 
$5.484ms$ ($5484 \mu s$)
due to the L-SIG (one of the legacy 
preamble's fields) duration limit~\cite{IEEEBase1}.
The A-MPDU frame structure in two-level aggregation
is shown in Figure~\ref{fig:twole}.

IEEE 802.11ax also enables extension of the acknowledgment 
mechanism by using
an acknowledgment window of 256 MPDUs.
In this paper 
we also assume that all MPDUs 
transmitted in an A-MPDU frame are 
from the same Traffic Stream (TS). 
In this case up to 256 MPDUs are
allowed in an A-MPDU frame of 11ax. 

Finally, in 11ax it 
is possible to transmit/receive simultaneously 
to/from up to 74 stations over the DL/UL 
respectively using MU. 

\section{Model}

\subsection{HE scheduling strategies for TCP Usage}

We compare between 11ax contention 
based Single User (SU), Reverse Direction (RD) SU and Multi
User (MU) TCP DL unidirectional scheduling strategies in order to optimize
the performance of DL single direction TCP connections,
from the AP to stations.

\subsubsection{Scheduling strategy 1 - HE DL 
Single User Reverse Direction unidirectional TCP}

Recall that Reverse Direction (RD) is a mechanism by which the
owner of a Transmission Opportunity (TXOP), the AP
in our case,
can enable its receiver to immediately transmit
back the TCP Acks during the TXOP so that the receiver does not need
to initiate UL transmission by using the
Extended Distributed Coordination
Function (EDCF) channel access method
defined in IEEE 802.11e~\cite{IEEEBase1}.
This is particularly efficient for
bi-directional traffic such as TCP Data/Ack segments
as it reduces overhead caused by collisions.

\begin{figure}
\vskip 15cm
\includegraphics{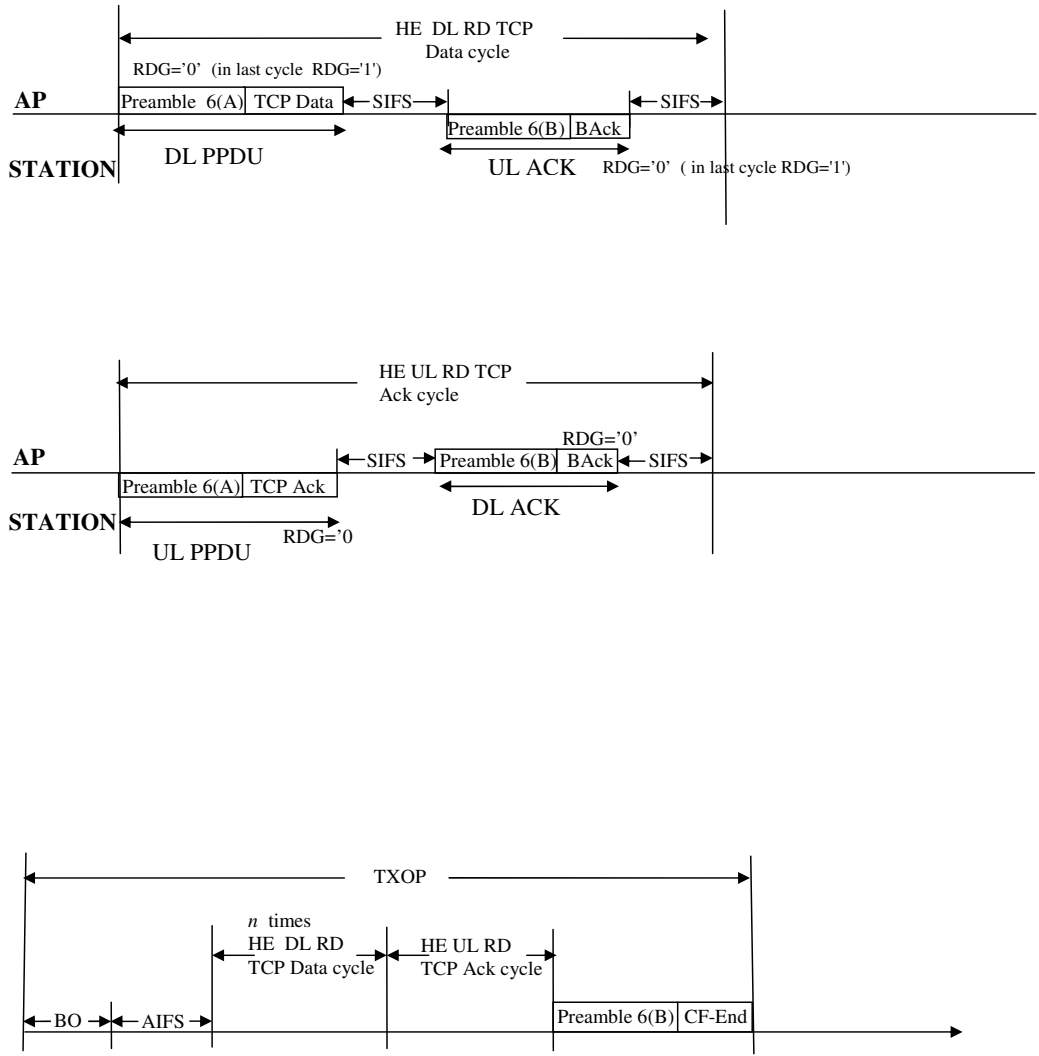}
\caption{Scheduling strategy 1: the scheduling strategy that uses HE Reverse Direction.}
\label{fig:rd}
\end{figure}

We examine a HE RD based scheduling strategy in which the AP transmits
DL HE  SU A-MPDU frames containing MPDUs of TCP Data segments
to a station and enables
the station to answer with an UL HE SU A-MPDU frame containing
MPDUs frames of
TCP Acks segments. 
Both the AP and the stations apply the two-level aggregation. 
We assume the following scenario to use
RD, as
is illustrated in Figure~\ref{fig:rd}.

After waiting AIFS and BackOff 
according to the 802.11 air access EDCA procedure,
the AP initiates a TXOP by transmitting 
$n$ DL HE SU A-MPDU frames in a row. 
Every such DL PPDU transmission, followed by receiving
the BAck frame from the station,
is denoted a {\it HE DL RD TCP Data cycle}.
In its last DL HE SU A-MPDU frame the
AP sets the RDG bit~\cite{IEEEBase1}, enabling the station
to respond with an UL HE SU A-MPDU frame
containing TCP Ack segments.
The AP then responds with a BAck frame
and terminates the TXOP with the CF-End frame~\cite{IEEEBase1}.
The transmission of the UL HE SU A-MPDU frame by the station,
followed by the BAck transmission from the AP, is denoted
a {\it HE UL RD TCP Ack cycle}.

In this HE RD based scheduling strategy we assume that there are
no collisions and TXOP are repeated over the channel
one after the other.
This is made possible by configuring the stations
in a way that prevents collisions.
For example, the stations
are configured 
to choose their BackOff intervals
from very large contention intervals, other than
the default ones~\cite{IEEEBase1}.  Thus, the
AP always wins over the channel without collisions.

In the case where the AP maintains TCP connections with
$S$ stations in parallel, it transmits to the stations
using Round Robin i.e. , after maintaining a TXOP with a station
the AP initiates a TXOP with the next station 
and so on.

\subsubsection{Scheduling strategy 2 - HE DL Single User contention based unidirectional TCP}

\begin{figure}
\vskip 11cm
\includegraphics{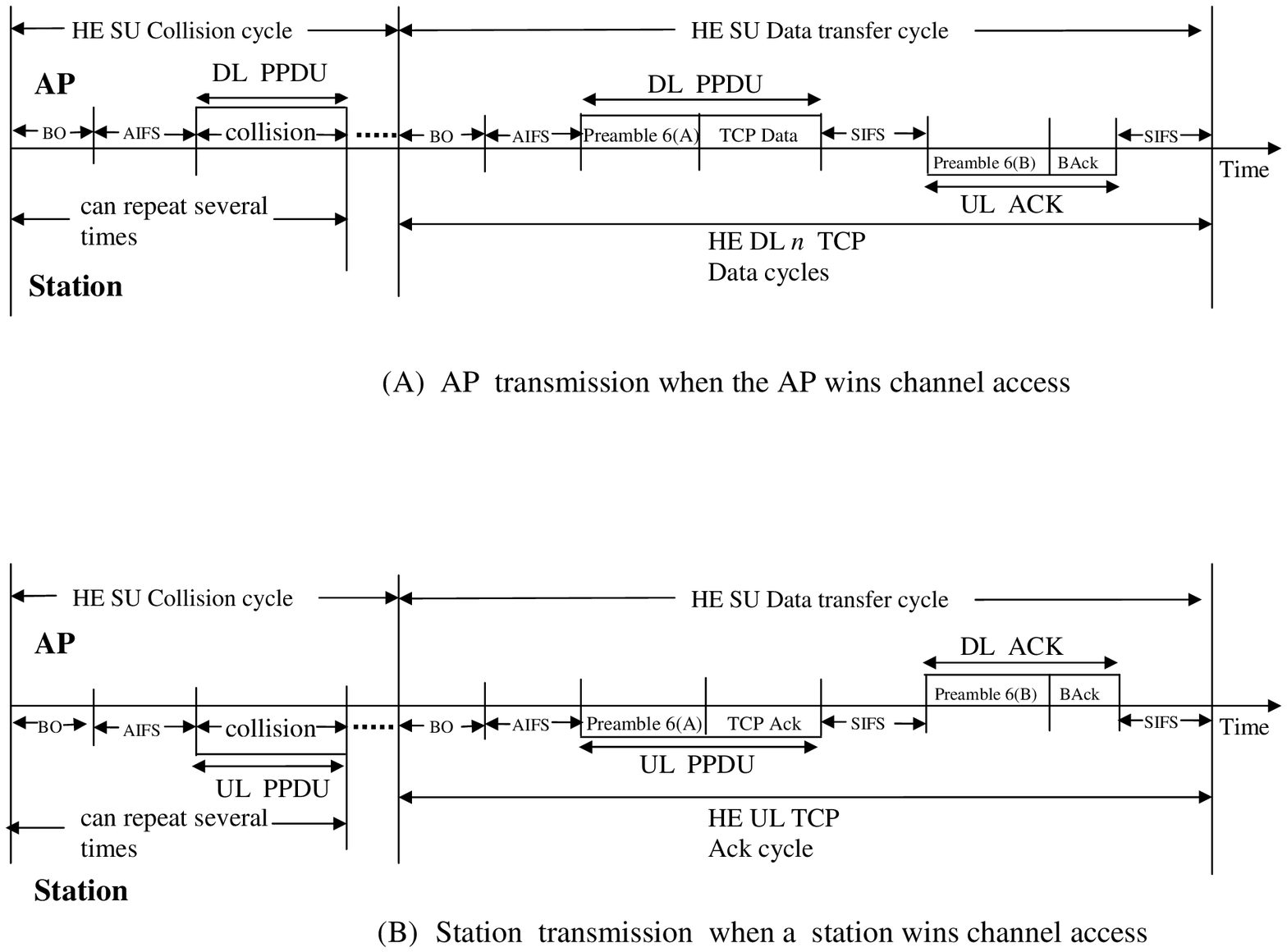}
\caption{Scheduling strategy 2: the contention based scheduling strategy .}
\label{fig:col}
\end{figure}

This HE SU scheduling strategy is 
shown in Figure~\ref{fig:col}.
In this strategy the AP uses TXOPs but
not RD: when the AP gets access to the
channel it transmits DL HE SU A-MPDU frames
containing TCP Data segments to a station
in a row. Every transmission of a single
DL HE SU A-MPDU frame from the AP is followed
by a BAck frame transmission from the
destination station; see Figure~\ref{fig:col}(A).

In this scheduling strategy both the AP and the stations contend in parallel
for accessing the air channel in every transmission
attempt, using the EDCF channel access method. 
In case the AP fails to gain
access to the channel during its first attempt, it tries
to access the channel
again according to EDCF, with re-try penalty (longer
BackOff interval) as shown in Figure~\ref{fig:col}(A).

The AP transmits to the stations in a Round Robin fashion.
After transmitting TCP Data segments to a station,
the AP does not serve that station again
before receiving TCP Ack segments from the station
and before the AP returns again to the station
in the Round Robin order. Notice from the above that
if the AP
returns to a station in the Round Robin order
before that station transmits TCP Ack segments
to the AP, the AP skips over the station.

A station transmits to the AP only when it has TCP
Ack segments, and it transmits the TCP Acks in one
UL HE SU A-MPDU frame. See Figure~\ref{fig:col}(B).





\subsubsection{Scheduling strategy 3 - HE DL 
simultaneous Multi User unidirectional TCP}

\begin{figure}
\vskip 15cm
\includegraphics{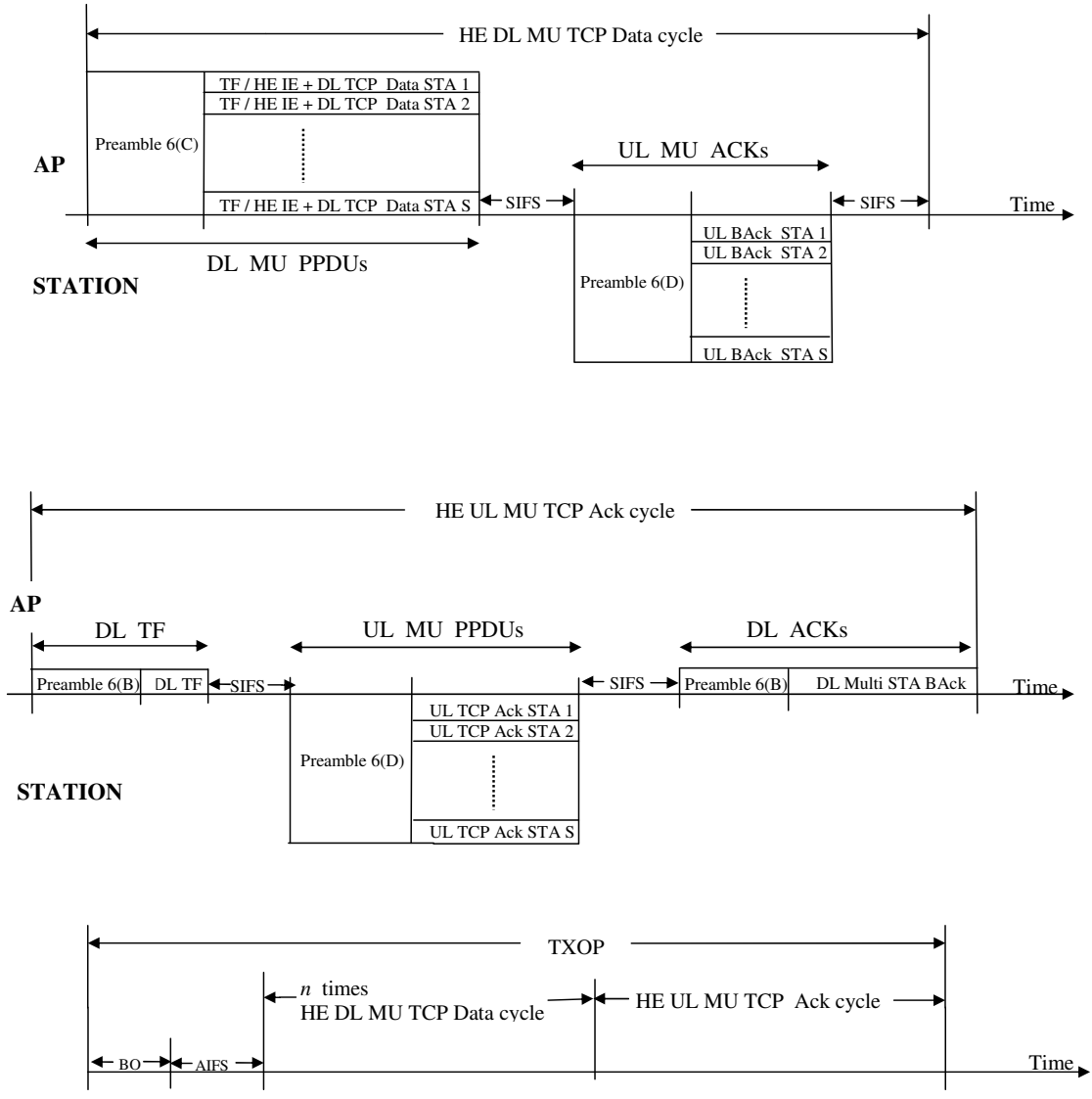}
\caption{Scheduling strategy 3: the HE Multi User scheduling strategy.}
\label{fig:mu}
\end{figure}

In the HE DL unidirectional TCP
Multi User mode the AP transmits TCP
Data to and receives TCP Acks
from several stations in parallel.
We assume the following
DL unidirectional TCP where simultaneous DL TCP Data
is sent by the AP to multiple stations
in the same PPDU and the TCP Acks are sent simultaneously by the
stations
at the same TXOP by using
Multi User, as
is illustrated in Figure~\ref{fig:mu}.

In this HE DL MU scheduling strategy,
after waiting
the BackOff and AIFS intervals, the AP
receives an air access and starts a TXOP by transmitting
$n$ DL HE MU A-MPDU frames
containing TCP Data segments to a group of
stations simultaneously. In every DL HE MU A-MPDU
frame the AP transmits to a different set of
stations in the group. After
receiving the UL HE MU BAck frames from the group
of stations simultaneously, the so-called
HE DL MU TCP Data cycle ends
and such a cycle
can now repeats itself several times. 

In order to transmit to a group of stations simultaneously,
the AP allocates Resource Units (RU), i.e.
sub-channels, per served station.
RU allocation is done at the DL for TCP Data segments
and at the UL for the TCP Acks.
The AP signals the stations when and how to transmit, i.e.
their UL RU allocation using
one of two possible methods.
In the first method
the AP transmits a
unicast Trigger
Frame (TF) to every station that contains the UL RU allocation. 
This frame is a control 
MPDU frame that is added to the other Data MPDUs
that the AP transmits to a station in a DL HE MU A-MPDU
frame.
The alternative
method is
to add an HE Control Element to $every$
MPDU in the DL HE MU
A-MPDU frame that is transmitted to every station.
In the following Goodput computations we optimize
the amount of overhead used due to the above methods
by computing the minimum overhead needed as a function
of the number of data MPDUs in the DL HE MU A-MPDU frame.

At the end of the last HE DL MU TCP Data cycle the AP
initiates a HE UL MU TCP Ack cycle by
transmitting the broadcast
Trigger Frame (TF). 
This TF solicits TCP Ack transmissions from the
stations to the AP.
At this transmission
the stations transmit TCP Ack segments
using UL HE MU A-MPDU frames. Every station transmits
its TCP Ack segments in a different UL HE MU A-MPDU frame.
The AP acknowledges the stations' UL HE MU A-MPDU
frames by generating and transmitting a single DL Multi Station BAck
frame. At this stage the HE UL MU TCP Ack cycle ends 
and a new series of HE DL MU TCP Data cycle(s) and
HE MU UL TCP Ack cycle begin.

As in the SU RD based
scheduling strategy we assume that
there are no collisions
by increasing
the size of the congestion interval
from which the stations choose their EDCF BackOff 
extended interval.

\subsection{IEEE 802.11 Frames/PPDU formats}

\begin{figure}
\vskip 15cm
\includegraphics{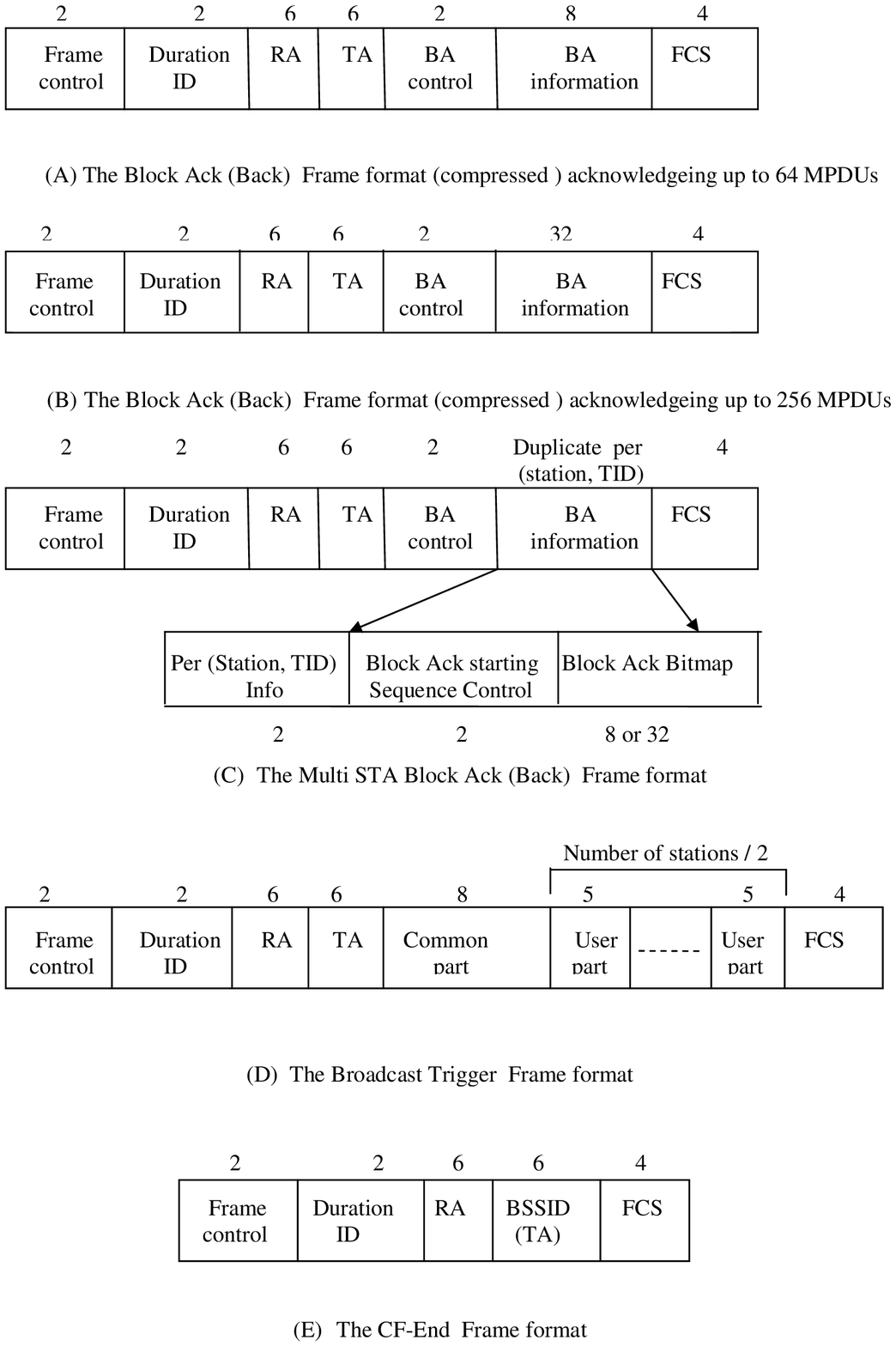}
\caption{The BAck, Multi Station BAck, Trigger Frame and CF-End frames' formats.}
\label{fig:ff}
\end{figure}

\begin{figure}
\vskip 15cm
\includegraphics{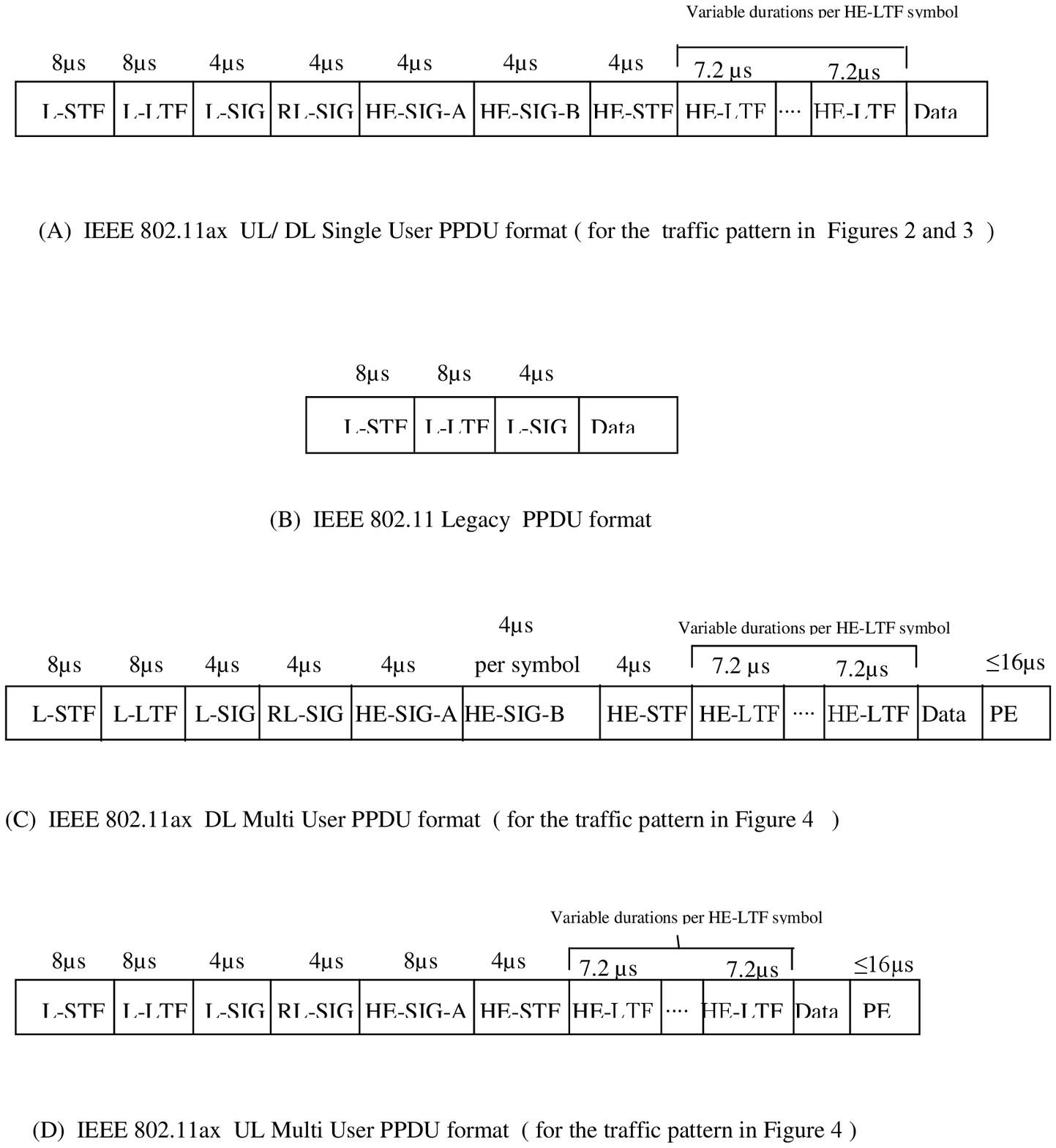}
\caption{The PPDUs' formats in the SU and MU modes.}
\label{fig:preamble}
\end{figure}

In Figure~\ref{fig:ff} we show the 802.11 frames' formats
of the BAck, Multi Station BAck, TF and CF-End frames 
used in the various scheduling strategies.
In Figure~\ref{fig:preamble} we show the various PPDUs' formats
used in the various scheduling strategies shown in
Figures~\ref{fig:rd}-\ref{fig:mu}. 

For the TCP Data/Ack segments' transmission in Figure~\ref{fig:rd},
scheduling strategy 1,
the PPDU format in Figure~\ref{fig:preamble}(A) is used
while the BAck and CF-End frames are transmitted using
the legacy mode in Figure~\ref{fig:preamble}(B).

For the TCP Data/Ack segments' transmission in Figure~\ref{fig:col},
scheduling strategy 2,
the PPDU in Figure~\ref{fig:preamble}(A) is used
while the BAck frames are transmitted by the legacy mode
shown in Figure~\ref{fig:preamble}(B).

For the TCP DL Data segments' transmission in Figure~\ref{fig:mu},
scheduling strategy 3,
the PPDU format in Figure~\ref{fig:preamble}(C) is used
and the BAcks are transmitted using the PPDU format
in Figure~\ref{fig:preamble}(D). The TCP UL Ack segments
are transmitted by the PPDU format in Figure~\ref{fig:preamble}(D)
and the TF and DL Multi Station Ack frames are transmitted
by the legacy PPDU format in Figure~\ref{fig:preamble}(B).

In the 11ax PPDU formats we find the HE-LTF fields,
the number of which equals the number of SSs
in use; 4 in our case.
In this paper we assume that each such field is
composed of 2X LTF and therefore of duration
$7.2 \mu s$~\cite{IEEEax}.

Notice also that the PSDU frame in 11ax contains
a Packet Extension (PE) field.
This field is mainly used in MU mode
and we assume that it is $0 \mu s$ in 
SU and the longest possible in MU,
$16 \mu s$.

In the HE-SIG-B field used in the PPDU format
of Figure~\ref{fig:preamble}(C) the Modulation/Coding Scheme (MCS) 
that is used
for this field is the minimum between MCS4 and the one used
for the data transmissions~\cite{IEEEax}. The length
of this field is also a function of the number of stations
to which the AP transmits simultaneously. Therefore,
in the case of 4 stations for example, the HE-SIG-B field
duration is $8 \mu s$ for MCS0 and MCS1 and $4 \mu s$
for MCS2-4 following section 23.3.9.8 in~\cite{IEEEax}. For
MCS5-MCS11 it is $4 \mu s$ as for  MCS4.

\newpage

\subsection{Parameters' values}

We assume the 5GHz band, a 160MHz channel and that the
AP and each station has 4 antennae.
In SU mode, i.e. in scheduling strategies 1 and 2,
the AP and the stations use up to 4 Spatial Streams and the entire
channel is devoted to transmissions of the AP
and stations. The BAck frames are transmitted
using legacy mode and the basic rates' set is used.
The PHY rate 
$R_{legacy}$ is set to the largest basic rate that
is smaller or equal to
the TCP Data/Ack segments' transmission rate $R_{TCP}$.

In Table~\ref{tab:phyratesSU} we show the
PHY rates and the length of preambles
used in SU mode in scheduling strategies 1 and 2 and
in the various MCSs. The values are taken from~\cite{IEEEax}.

\begin{table}
\caption{\label{tab:phyratesSU}{The PHY rates and preambles
in IEEE 802.11ax used in Single User mode and in scheduling strategies 1 and 2. 
A 160 MHz channel is assumed, with 4 Spatial Streams. The BAck frames are conducted at the basic rates' set.}} 
\vspace{3 mm}
\tiny
\center
\begin{tabular}{|r|c|c|c|c|c|}  \hline
     & \multicolumn{2}{|c|}1 &  &
       \multicolumn{2}{|c|}2  \\ \cline{2-6}
     & \multicolumn{2}{c|}{SU UL/DL data}&  &
       \multicolumn{2}{c|}{SU UL/DL BAck } \\
     & \multicolumn{2}{c|}{transmission rate }  &  &
       \multicolumn{2}{c|}{transmission rate } \\ \hline
     &  PHY Rate & Preamble & 
        & PHY Rate (legacy) & Preamble \\
MCS  &  (Mbps)   & ($\mu s$) 
     &  & (Mbps)   & ($\mu s$) \\
     &  GI$=0.8 \mu s$   &
       & & GI$=0.8 \mu s$ & \\ \hline
      \multicolumn{6}{|c|}{}  \\ \hline
 0   &  288.2  & 64.8  &   & 48.0 & 20.0   \\ 
 1   &  576.5  & 64.8  &   & 48.0 & 20.0   \\ 
 2   &  864.7  & 64.8  &   & 48.0 & 20.0   \\ 
 3   & 1152.9  & 64.8  &   & 48.0 & 20.0   \\ 
 4   & 1729.4  & 64.8  &   & 48.0 & 20.0   \\ 
 5   & 2305.9  & 64.8  &   & 48.0 & 20.0   \\ 
 6   & 2594.1  & 64.8  &   & 48.0 & 20.0   \\ 
 7   & 2882.4  & 64.8  &   & 48.0 & 20.0   \\ 
 8   & 3458.8  & 64.8  &   & 48.0 & 20.0   \\ 
 9   & 3848.1  & 64.8  &   & 48.0 & 20.0   \\ 
10   & 4323.5  & 64.8  &   & 48.0 & 20.0   \\ 
11   & 4803.9  & 64.8  &   & 48.0 & 20.0   \\ \hline 
\end{tabular}  
\end{table}

When using MU mode in scheduling strategy 3,
the 160MHz channel is divided into $\frac{S}{4}$ channels
of $\frac{160 \cdot 4}{S}$ MHz each, $S=4, 8, 16, 32, 64$.
When $S=4$ the 160MHz is used in MU-MIMO.
For $S>4$ MU-MIMO+OFDMA is used. 
The AP transmits over the DL 
to 4 stations in every such channel, by allocating
a single Spatial Stream per station i.e. in every channel
4 Spatial Streams are allocated. 
For example, for $S=64$ there are 16 channels
of 10MHz each; in each one the AP transmits to 4 stations.
The stations transmit
to the AP over the UL in a symmetrical way to that of the
AP over the DL. 

The TF and the Multi Station BAck frames are
transmitted using the legacy mode and the
PHY rate 
$R_{legacy}$ is set to the largest basic rate that
is smaller or equal to
the TCP Data/Ack segments' transmission rate $R_{TCP}$.
The minimal basic PHY rate is 6Mbps. In the case
of $R_{TCP}$ smaller than 6Mbps, $R_{legacy}$ is never less
than 6Mbps. This can occur in the case of 64 stations.

In Table~\ref{tab:phyrates} we show the PHY rates and the preambles
used in scheduling strategy 3, in the various MCSs and
in all cases of the number of stations $S$, i.e. 
$S= 4, 8, 16, 32$ and 64.

\begin{table}
\caption{\label{tab:phyrates}{The PHY rates and preambles in IEEE 802.11ax in scheduling strategy 3. A 160 MHz channel is assumed, with 4 Spatial Streams. The TF and BAck
transmissions are conducted at the basic rate set. }}
\vspace{3 mm}
\tiny
\center
\begin{tabular}{|r|c|c|c|c|c|c|c|}  \hline
     & \multicolumn{2}{|c|}1  & \multicolumn{2}{|c|}2 & &
       \multicolumn{2}{|c|}3  \\ \cline{2-8}
     & \multicolumn{2}{c|}{MU UL data}& \multicolumn{2}{c|}{MU DL data}&  & \multicolumn{2}{c|}{DL TF/Multi Station BAck}  \\ 
& \multicolumn{2}{c|}{transmission rate} & \multicolumn{2}{c|}{transmission rate} &  & \multicolumn{2}{c|}{transmission rate}  \\ \hline
     &  PHY Rate & Preamble & PHY Rate & Preamble & 
        & PHY Rate (legacy) & Preamble \\
MCS  &  (Mbps per 1 SS)   & ($\mu s$)  & (Mbps per 1 SS) & ($\mu s$)
     & & (Mbps)   & ($\mu s$) \\
     &  GI$=1.6 \mu s$   & & GI$=0.8 \mu s$ &
      &  & GI$=0.8 \mu s$ &  \\ \hline
      \multicolumn{8}{|c|}{4 stations}  \\ \hline
 0   &  68.1   & 64.8 & 72.1 & 72.8 && 48.0 & 20.0 \\ 
 1   &  136.1  & 64.8 &144.1 & 72.8 && 48.0 & 20.0 \\ 
 2   &  204.2  & 64.8 &216.2 & 68.8 && 48.0 & 20.0 \\ 
 3   &  272.2  & 64.8 &288.2 & 68.8 && 48.0 & 20.0 \\ 
 4   &  408.3  & 64.8 &432.4 & 68.8 && 48.0 & 20.0 \\ 
 5   &  544.4  & 64.8 &576.5 & 68.8 && 48.0 & 20.0 \\ 
 6   &  612.5  & 64.8 &648.5 & 68.8 && 48.0 & 20.0 \\ 
 7   &  680.6  & 64.8 &720.6 & 68.8 && 48.0 & 20.0 \\ 
8    &  816.7  & 64.8 &864.7 & 68.8 && 48.0 & 20.0 \\ 
 9   &  907.4  & 64.8 &960.7 & 68.8 && 48.0 & 20.0 \\ 
10   & 1020.8  & 64.8 &1080.4 & 68.8 && 48.0 & 20.0 \\ 
11   & 1134.2  & 64.8 &1201.0 & 68.8 && 48.0 & 20.0 \\ \hline 
      \multicolumn{8}{|c|}{8 stations}  \\ \hline
 0   &  34.0 & 64.8 &36.0 & 76.8 && 36.0 & 20.0  \\ 
 1   &  68.1 & 64.8 &72.1 & 76.8 && 48.0 & 20.0  \\ 
 2   & 102.1 & 64.8 &108.1 & 72.8 && 48.0 & 20.0  \\ 
 3   & 136.1 & 64.8 &144.1 & 72.8 && 48.0 & 20.0  \\ 
 4   & 204.2 & 64.8 &216.2 & 68.8 && 48.0 & 20.0  \\ 
 5   & 272.2 & 64.8 &288.2 & 68.8 && 48.0 & 20.0  \\ 
 6   & 306.3 & 64.8 &324.3 & 68.8 && 48.0 & 20.0  \\ 
 7   & 340.3 & 64.8 &360.3 & 68.8 && 48.0 & 20.0  \\ 
8    & 408.3 & 64.8 &432.4 & 68.8 && 48.0 & 20.0  \\ 
 9   & 453.7 & 64.8 &480.4 & 68.8 && 48.0 & 20.0  \\ 
10   & 510.4 & 64.8 &540.4 & 68.8 && 48.0 & 20.0  \\
11   & 567.1 & 64.8 &600.4 & 68.8 && 48.0 & 20.0  \\ \hline 
     \multicolumn{8}{|c|}{16 stations}  \\ \hline
 0   &  16.3 & 64.8 &17.2 & 84.8 && 12.0 & 20.0 \\ 
 1   &  32.5 & 64.8 &34.4 & 84.8 && 12.0 & 20.0 \\ 
 2   &  48.8 & 64.8 &51.6 & 76.8 && 24.0 & 20.0 \\ 
 3   &  65.0 & 64.8 &68.8 & 76.8 && 48.0 & 20.0 \\ 
 4   &  97.5 & 64.8 &103.2 & 72.8 && 48.0 & 20.0 \\ 
 5   & 130.0 & 64.8 &137.6 & 72.8 && 48.0 & 20.0 \\ 
 6   & 146.3 & 64.8 &154.9 & 72.8 && 48.0 & 20.0 \\ 
 7   & 162.5 & 64.8 &172.1 & 72.8 && 48.0 & 20.0 \\ 
8    & 195.0 & 64.8 &206.5 & 72.8 && 48.0 & 20.0 \\ 
 9   & 216.7 & 64.8 &229.4 & 72.8 && 48.0 & 20.0 \\ 
10   & 243.8 & 64.8 &258.1 & 72.8 && 48.0 & 20.0 \\ 
11   & 270.8 & 64.8 &286.8 & 72.8 && 48.0 & 20.0 \\ \hline 
\end{tabular} 
\end{table}

\addtocounter{table}{-1}

\begin{table}
\caption{\label{tab:phyrates1}{(cont.)}}
\vspace{3 mm}
\tiny
\center
\begin{tabular}{|r|c|c|c|c|c|c|c|}  \hline
     & \multicolumn{2}{|c|}1  & \multicolumn{2}{|c|}2 & &
       \multicolumn{2}{|c|}3  \\ \cline{2-8}
     & \multicolumn{2}{c|}{MU UL data}& \multicolumn{2}{c|}{MU DL data}&  & \multicolumn{2}{c|}{DL TF/Multi Station BAck}  \\ 
& \multicolumn{2}{c|}{transmission rate} & \multicolumn{2}{c|}{transmission rate} &  & \multicolumn{2}{c|}{transmission rate}  \\ \hline
     &  PHY Rate & Preamble & PHY Rate & Preamble & 
        & PHY Rate (legacy) & Preamble \\
MCS  &  (Mbps per 1 SS)   & ($\mu s$)  & (Mbps per 1 SS) & ($\mu s$)
     & & (Mbps)   & ($\mu s$) \\
     & GI$=1.6 \mu s$ & & GI$=0.8 \mu s$ &
      &  & GI$=0.8 \mu s$ &  \\ \hline
     \multicolumn{8}{|c|}{32 stations} \\ \hline
 0   &   8.1 & 64.8 &8.6& 104.8 &&  6.0 & 20.0 \\ 
 1   &  16.3 & 64.8 &17.2& 104.8 && 12.0 & 20.0 \\ 
 2   &  24.4 & 64.8 &25.8& 84.8 && 24.0 & 20.0 \\ 
 3   &  32.5 & 64.8 &34.4& 84.8 && 24.0 & 20.0 \\ 
 4   &  48.8 & 64.8 &51.6& 80.8 && 48.0 & 20.0 \\ 
 5   &  65.0 & 64.8 &68.8& 80.8 && 48.0 & 20.0 \\ 
 6   &  73.1 & 64.8 &77.4& 80.8 && 48.0 & 20.0 \\ 
 7   &  81.3 & 64.8 &86.0& 80.8 && 48.0 & 20.0 \\ 
8    &  97.5 & 64.8 &103.2& 80.8 && 48.0 & 20.0 \\ 
 9   & 108.3 & 64.8 &114.7& 80.8 && 48.0 & 20.0 \\ 
10   & 121.9 & 64.8 &129.0& 80.8 && 48.0 & 20.0 \\ 
11   & 135.4 & 64.8 &143.4& 80.8 && 48.0 & 20.0 \\ \hline 
     \multicolumn{8}{|c|}{64 stations}  \\ \hline
 0   &  3.5 & 64.8 &3.8& 136.8 &&  6.0 & 20.0 \\ 
 1   &  7.1 & 64.8 &7.5& 136.8 &&  6.0 & 20.0 \\ 
 2   & 10.6 & 64.8 &11.3& 100.8 &&  9.0 & 20.0 \\ 
 3   & 14.2 & 64.8 &15.0& 100.8 && 12.0 & 20.0 \\ 
 4   & 21.3 & 64.8 &22.5& 88.8 && 18.0 & 20.0 \\ 
 5   & 28.3 & 64.8 &30.0& 88.8 && 24.0 & 20.0 \\ 
 6   & 31.9 & 64.8 &33.8& 88.8 && 24.0 & 20.0 \\ 
 7   & 35.4 & 64.8 &37.5& 88.8 && 24.0 & 20.0 \\ 
 8   & 42.5 & 64.8 &45.0& 88.8 && 36.0 & 20.0 \\ 
 9   & 47.2 & 64.8 &50.0& 88.8 && 36.0 & 20.0 \\ 
10   &  N/A &  N/A &N/A& N/A &&  N/A &  N/A \\ 
11   &  N/A &  N/A &N/A& N/A &&  N/A &  N/A \\ \hline 
\end{tabular}  
\end{table}

We assume the
Best Effort Access Category
in which $AIFS=43 \mu s$ for
the AP and $52 \mu s$ for a station, $SIFS=16 \mu s$ and $CW_{min}=16$
for the transmissions of the AP.
Recall that in scheduling strategies
1 and 3 we assume there are
no collisions between the AP and the stations.
The BackOff interval is a random number
chosen uniformly from
the range $[0,....,CW_{min}-1]$. 
Since we consider a very 'great' number
of transmissions from the AP in scheduling
strategies 1 and 3, we take the BackOff average
value of  
$\ceil{\frac{CW_{min}-1}{2}}$, and the average BackOff interval 
for the AP is
$\ceil{\frac{CW_{min}-1}{2}} \cdot SlotTime$
which equals $67.5 \mu s$ for a $SlotTime= 9 \mu s$.

Concerning the transmission in non-legacy mode,
an OFDM symbol
is $12.8 \mu s$. In the DL
we assume a GI of $0.8 \mu s$ and therefore the symbol
in this direction is $13.6 \mu s$. In the UL MU we assume
a GI of $1.6 \mu s$ and therefore the symbol in this
direction is $14.4 \mu s$. The UL GI is $1.6 \mu s$ 
due to UL arrival time variants. In UL SU the GI is $0.8 \mu s$.
When considering transmissions in legacy mode, the symbol
is $4 \mu s$ containing a GI of $0.8 \mu s$.

We assume that the MAC Header field
is of 28 bytes and
the Frame Control Sequence (FCS) field is of 4 bytes.
Finally, we assume
TCP Data segments of $L_{DATA}=1460, 464$ and
$208$ bytes. Therefore, the resulting
MSDUs' lengths are $L^{'}_{DATA}=1508, 512$ and $256$
bytes respectively ( 20 bytes of TCP header
plus 20 bytes of IP header plus
8 bytes of LLC SNAP are added ).
Together with the SubHeader field and rounding
to an integral multiple of 4 bytes, every MSDU is now
of $Len^{D}=1524, 528$ and $256$ bytes respectively. 
Due to the limit
of 11454 bytes on the MPDU size, 7, 21
and 42 such MSDUs are possible respectively in one MPDU.

The TCP receiver transmits TCP Acks. Every MSDU containing
a TCP Ack
is of $L^{'}_{Ack}=48$ bytes 
( 20 bytes of TCP Header + 20 bytes of IP header +
8 bytes of LLC SNAP ). Adding 14 bytes of the SubHeader
field and rounding to an integral multiple of 4 bytes, every
MSDU is of $Len^{A}=64$ bytes, and every
single MPDU, again due to the size limit
of 11454 bytes, can contain up to 178 MSDUs (TCP Acks).
Thus, the receiver can transmit
up to $N_{MAX}=256 \cdot 178$ TCP
Acks in a single HE UL A-MPDU frame. 

\section{Goodput analysis}

The system Goodput analysis has two targets.
The first is to find the optimal working
point for  each of the proposed scheduling strategies, i.e. finding the
working point that maximizes the Goodput of a TXOP.
By an optimal working point we refer to the number
of DL TCP Data MSDUs to transmit in a TXOP, how many HE DL TCP Data cycles
to transmit in a TXOP, and the
optimal HE DL A-MPDU structure within
each HE DL TCP Data cycle i.e. the number of MPDUs in the 
HE DL A-MPDU and
the number of MSDUs within every MPDU.

Notice that the Goodput computed is actually the 
system {\it TCP Goodput},
i.e. the average number of TCP Data bits that are transmitted
in the system per time unit. However, in SU,
scheduling strategies 1 and 2, when
$S$ stations are served by the AP, a single station enjoys a given TCP
Googput $G$ in every $S^{th}$ TXOP only. 
The system provides a TCP Goodput $G$ to {\it all} the stations
over $S$ TXOPs.
In the MU strategy, scheduling strategy 3, the TCP Goodput
$G$ of a TXOP is that provided to all the stations together 
over one TXOP.

The second
target of the analysis is to find the time intervals
over which the system enables a given TCP Goodput $G$ to all of its
stations. A scheduling strategy that enables a given TCP Goodput
to all stations over shorter time intervals is more efficient.

\subsection{Maximum Goodput of a TXOP}

\noindent
Computing the optimal working point per scheduling strategy, i.e. the
one that maximizes the Goodput of a TXOP, is done in 3 stages:

\begin{enumerate}

\item
The number of TCP Data segments that can be transmitted
in a TXOP is limited by the number $N_{MAX} = 256 \cdot 178$ TCP Ack
segments that can be transmitted in one HE UL A-MPDU frame.
The number 256 is the Block Ack window size and 178 is
the number of TCP Ack segments possible in one MPDU.
$N_{MAX}$ was computed in Section 3.3 and
if more than $N_{MAX}$ TCP Data segments are transmitted
in a TXOP, stations begin to accumulate TCP Acks and
the Goodput {\it decreases}.

\item
In order to maximize the Goodput the TCP Acks shall be transmitted
in the minimal possible number of MPDUs in order to minimize
overhead associated with MPDUs containing TCP Acks.
For $N$ TCP Acks, $1 \le N \le N_{MAX}$ the number
of MPDUs is $\ceil{\frac{N}{178}}$. See Section 3.3 .

\item
For every number $N$ of TCP Data segments transmitted
in a TXOP, $1 \le N \le N_{MAX}$, it is necessary to determine
what is the number of HE DL A-MPDU frames for transmission in a TXOP, and 
the structure of each HE DL A-MPDU frame, i.e. how many MPDUs
are in every HE DL A-MPDU frame and how many MSDUs are
in every MPDU. This is necessary in order to minimize
overhead associated with MPDU and HE DL A-MPDU frames that
contain TCP Data segments. This computation is carried on in the appendix.
\end{enumerate}

For scheduling strategies 1 and 3 we provide
a mathematical analysis to compute the Goodput of a TXOP.
This analysis was verified by the NS3 simulator.
The analysis and simulation results
match perfectly. This is not surprising
as there is no 
stochastic process in these
strategies. Therefore, we later omit
their simulation results.
For scheduling strategy 2 we only use simulation
to compute the Goodput.

\subsubsection{Goodput analysis for scheduling strategy 1 - HE DL Single User Reverse Direction unidirectional TCP}

For this HE SU RD scheduling strategy
notice that $N_{MAX}=256 \cdot 178$
TCP Acks that a station can transmit in one HE UL RD TCP Ack
cycle is an 
upper bound on the number of DL TCP Data MSDUs that
can be transmitted by the AP in a TXOP. Using a larger number
will cause the receiver to accumulate TCP Acks and the
Goodput to decrease. See Section 3.3 .


An HE SU RD TXOP has a fixed overhead of the AIFS 
and BackOff intervals,
and the transmission time of the CF-End 
frame and its associated preamble 6(B).
In addition there are overheads associated with the 
transmission of an MPDU frame and a HE DL SU A-MPDU frame.
The MPDU's overhead is composed of the
MAC Header, MPDU Delimiter and the FCS fields. The 
HE DL SU A-MPDU overhead for scheduling 
strategy 1 is $Pr(6(A))+Pr(6(B))+T(BAck)
+2 \cdot SIFS$. See Figure~\ref{fig:rd} .

Due to the various overheads
it may 
not always be worthwhile to transmit
$N_{MAX}$ TCP Data MSDUs in a TXOP.
For instance, a single MSDU
can cause the creation of both a new HE DL SU A-MPDU and an MPDU with
all of 
the associated overhead that overall can reduce the
Goodput.

Assume that $N$ TCP Data MSDUs are transmitted in a TXOP.
For a reliable clearly it is most efficient to
transmit the $N$ TCP Ack MSDUs in the smallest number of MPDUs,
i.e. in $\ceil{\frac{N}{178}}$ MPDUs.
This will reduce the overhead associated with MPDUs containing
TCP Ack MSDUs to a minimum.
Recall that all the TCP Ack MSDUs are transmitted in one
HE UL SU A-MPDU frame.

Recall now that in the appendix we show how to schedule $N$ TCP Data MSDUs
in the most efficient way in a TXOP,
i.e. the scheduling that results in the smallest
HE UL SU A-MPDUs' and MPDUs' overheads.
Assume that the optimal scheduling of $N$ TCP Data MSDUs
is in $n$ HE DL SU A-MPDUs. Let $A_i$ and $N_i$ be the numbers of MPDUs
and MSDUs respectively
in HE DL SU A-MPDU number $i,  1 \le i \le n$.

Let $AIFS$ and $BO$ denote the length,
in $\mu s$, of the AIFS and average BackOff
intervals. Let $O_M$ be the total length, in bytes, 
of the MAC Header, MPDU Delimiter and FCS fields.
The Goodput of
scheduling strategy 1
is given by
Eq.~\ref{equ:ss2}, assuming the AP transmits $N$ TCP Data MSDUs in
$n$ HE DL SU RD TCP Data cycles:

\begin{equation}
Goodput=
\frac
{N \cdot (L_{DATA} \cdot 8) }
{AIFS+BO(average)+ \sum_{i=1}^{n} T_{Data}^{Cycle}(A_i, N_i) + T_{Ack}^{Cycle}+
Pr(6(B))+T(CF-End) }
\label{equ:ss2}
\end{equation}

\normalsize

\noindent
Where:

\small

\begin{equation}
T_{Data}^{cycle}(A_i,N_i) = 
Pr(6(A))+TSym_{DL} \cdot 
\ceil{\frac{(N_i \cdot Len^{D} + A_i \cdot O_M) \cdot 8 + 22}{TSym_{DL} \cdot R_{DL}}} 
+Pr(6(B)) + T(BAck) + 2 \cdot SIFS
\\ \nonumber
\end{equation}
\begin{equation}
T_{Ack}^{cycle} =  Pr(6(A))+TSym_{UL} \cdot \ceil{\frac{(N \cdot Len^A + 
\ceil{\frac{N}{178}} \cdot O_M) 
\cdot 8 + 22}{TSym_{UL} \cdot R_{UL}}} +Pr(6(B)) + T(BAck) + 2 \cdot SIFS 
\\ \nonumber
\end{equation}

\begin{equation}
T(BAck) = TSym_{leg} \cdot \ceil{\frac{(54 \cdot 8) +22}{TSym_{leg} \cdot R_{leg}}} \\ \nonumber
\end{equation}
\begin{equation}
T(CF-End) = TSym_{leg} \cdot \ceil{(\frac{(20 \cdot 8) 
+22}{TSym_{leg} \cdot R_{leg}}} 
\end{equation}

\normalsize

$T(BAck)$ and $T(CF-End)$ are the transmission
times of the BAck and the CF-End frames respectively. 
The times are based on the frames' lengths shown in
Figure~\ref{fig:ff}. $T(BAck)$ assumes the BAck frame
acknowledging 256 MPDUs, Figure~\ref{fig:ff}(B).
If the number of MPDUs is
smaller than or equal to 64, the BAck
frame of Figure~\ref{fig:ff}(A) is used, and in this case
the term 54 in the numerator of $T(BAck)$ is replaced
by 30.

$TSym_{DL}$ and $TSym_{UL}$ are the lengths
of the OFDM symbols in the DL and UL respectively, including the GI,
when transmitting in a non-legacy mode,
and every transmission
must be of an integral number of OFDM symbols.
$TSym_{leg}=4 \mu s$  and it is the length of the OFDM
symbol that is used in legacy transmissions.
$R_{DL}$ and $R_{UL}$ are
the PHY rates for TCP Data/Ack transmissions on
the DL and the UL respectively, and $R_{leg}$
is the legacy PHY rate used for the transmission of the BAck
and CF-End frames. See Table~\ref{tab:phyratesSU}.
The additional 22 bits in the various numerators of
the frames' transmission times
are due to the SERVICE and TAIL fields that are added to every
transmission by the PHY layer conv. protocol~\cite{IEEEBase1}.

\subsubsection{Goodput analysis for Scheduling strategy 2 - HE DL Single User contention based unidirectional TCP}

In this scheduling strategy the AP
transmits $N$ DL TCP Data segments, $1 \le N \le N_{MAX}$,
to a station 
after accessing the channel. 
The number of HE DL SU A-MPDU frames containing these $N$
TCP Data segments and their structure is determined
as in scheduling strategy 1.
The AP does not
transmit to a station again before
it receives UL TCP Acks from the station 
and transmits to the stations in Round Robin fashion.
We check the performance of this scheduling strategy
for various values of $N$, the number of DL TCP Data segments that
the AP transmits in one transmission to the station, $1 \le N \le N_{MAX}$.


\subsubsection{Goodput analysis for Scheduling strategy 3 - HE DL simultaneous Multi User unidirectional TCP}

In scheduling strategy 3
every single DL
TCP connection between the AP and a station 
can be considered as the TCP connection in scheduling
strategy 1. 
The only difference is that in scheduling
strategy 3 the AP can transmit in several DL TCP connections
parallel to several stations, and
several stations can transmit their BAck and UL  TCP Ack segments
parallel to the AP. 
The analysis in
scheduling strategy 3 is therefore
basically the same as for scheduling strategy 1
with some differences specified below.


The Goodput of
scheduling strategy 3 shown in Figure~\ref{fig:mu}
is given by
Eq.~\ref{equ:ss4}, assuming the AP transmits $N$ DL TCP Data MSDUs in 
every TCP connection in $n$ HE DL MU A-MPDUs:


\begin{equation}
Goodput=
\frac
{N \cdot (L_{DATA} \cdot 8) \cdot S }
{AIFS+BO(Variable)+ \sum_{i=1}^{n} \cdot T_{Data}^{cycle}(A_i,N_i)+T_{Ack}^{cycle}}
\label{equ:ss4}
\end{equation}

\normalsize

\noindent
where:

\tiny

\begin{equation}
T_{Data}^{cycle}(A_i,N_i) =  Pr(6(C))+TSym_{DL} \cdot \ceil{\frac{(N_i \cdot Len^D + A_i \cdot ( O_M + 4)) \cdot 8+ 22}{TSym_{DL} \cdot R_{DL}}} +Pr(6(D)) + T(BAck) + 2 \cdot SIFS
\\ \nonumber
\end{equation}
\begin{equation}
T_{Ack}^{cycle} =  Pr(6(B))+T(TF)+Pr(6(D))+TSym_{UL} \cdot 
\ceil{\frac{(N \cdot Len^A + \ceil{\frac{N}{178}} \cdot O_M) \cdot 8 + 22}{TSym_{UL} \cdot R_{UL}}} +Pr(6(B)) + T(Mul.BAck) + 2 \cdot SIFS 
\\ \nonumber
\end{equation}

\begin{equation}
T(BAck) = TSym_{UL} \cdot \ceil{\frac{(54 \cdot 8) +22}{TSym_{UL} \cdot R_{UL}}} \\  \nonumber
\label{equ:ss3}
\end{equation}
\begin{equation}
T(TF) = TSym_{leg} \cdot \ceil{\frac{(28+(\frac{S}{2} \cdot 5) \cdot 8) +22}{TSym_{leg} \cdot R_{leg}}} 
\\ \nonumber
\end{equation}
\begin{equation}
T(Mul.BAck) = TSym_{leg} \cdot \ceil{\frac{((22+S \cdot 36) \cdot 8) +22}{TSym_{leg} \cdot R_{leg}}}
\\ 
\end{equation}

\normalsize

\indent
The quantity 4 in the numerator of the second
term in $T_{Data}^{cycle}(A_i,N_i)$ stands
for the HE IE added to every MPDU in order to schedule
parallel transmissions of the BAck frames
from the stations on the UL.
This holds for $1 \le A_i \le 18$. For $A_i > 18$ it 
is more efficient to contain a unicast TF frame of length 72 bytes,
containing the unicast TF frame (33 bytes)
and the MAC Header, MPDU Delimiter and FCS fields,
and rounding to an integral multiple of 4 bytes.

$T(BAck)$, $T(TF)$ and $T(Mul.BAck)$ are the transmission times
of the BAck, multicast TF and the Multi Station 
BAck frames respectively.
$T(MUl.BAck)$
is based on the Multi Station BAck frame
length given in Figure~\ref{fig:ff}(C) assuming
the acknowledgment of 256 MPDUs per HE DL A-MPDU frame.
When considering the acknowledgment of 64 MPDUs
the term 36 in the numerator is replaced by 12.
The term $S$ in $T(TF)$ and $T(Mul. BAck)$ denote
the number $S$ of stations transmitting data simultaneously
over the UL.

$TSym_{UL}$ and $TSym_{DL}$ are the lengths
of the OFDM symbols, containing the GIs,
used over the UL and DL respectively when transmitting
in non-legacy mode,
and every transmission
must be of an integral number of OFDM symbols.
$TSym_{leg}=4 \mu s$ and it is the length of the OFDM
symbol that is used in legacy transmissions.
$R_{DL}$ and $R_{UL}$ are the DL and UL PHY rates respectively
used for the transmission of TCP Data and TCP Acks
segments.
$R_{leg}$ is the legacy PHY rate used for the
transmission of the TF and Multi Station BAck frames.
The additional 22 bits in the various numerators of
the frames' transmission times
are due to the SERVICE and TAIL fields that are added to every
transmission by the PHY layer conv. protocol~\cite{IEEEBase1}.
 
\subsection{Goodput vs. delay computation}

For scheduling strategies 1 and 3 we 
measured the Goodput received in every TXOP
according to Eqs. 2 and 4 respectively.
In these equations the total number of
TCP Data bits transmitted in a TXOP is divided
by the TXOP length, measured in seconds. However,
since we assume that the same TXOPs repeat themselves one after
another, the computed Goodput of a TXOP is also
the Goodput of the system.

We now measure 
for every number $N$
of TCP Data segments 
transmitted in a TXOP, 
$1 \le N  \le N_{MAX}$
the resulted length of the TXOP interval containing the $N$ TCP
Data segments' and as
mentioned the Goodput is computed using
Eqs. 2 and 4 respectively.

For scheduling strategy 2 
we also measure the Goodput when transmitting 
$N$ TCP Data segments. However, in this scheduling
strategy 
there is no TXOP with RD and instead we measure
the average time elapsed from
the time the AP transmits to a station
TCP Data segments until it receives TCP Acks from the station.

From now on we denote by {\it cycle} the TXOPs
in scheduling strategies 1 and 3, and the above time
interval that we described for scheduling strategy 2.
By {\it cycle length} we denote the length, in seconds,
of the cycle.

The next step is as follows: Notice that for every
number $N$ of TCP Data segments transmitted in a cycle,
there is a resulting cycle length which shows how much time
is needed in a cycle for the transmission of these
TCP Data segments to a specific station. Thus, for every number $N$ of TCP Data
segments, $1 \le N \le N_{MAX}$ we attach two measures: the
cycle length in which these $N$ TCP Data segments are transmitted
and the resulting Goodput. 
We now arrange the cycle lengths in a list together with
the associated Goodputs in increasing order
of the cycle lengths.

Notice that two different cycle lengths can have the same Goodput.
One of the cycles has more TCP Data segments
but it can also have more A-MPDU/MPDUs' overhead.
IN addition, the number of TCP Data segments  can be large enough
so that the addition of one more  TCP Data segment barely changes
the Goodput.
For a set of  cycle lengths with the same Goodput we leave
only the shortest cycle in the list.

Consider now a cycle length of {\it L ms} with a Goodput $G$.
In scheduling strategies 1 and 2 (the 
SU ones) when the AP is communicating with $S$ stations
in Round Robin, a station receives TCP Data segments
in every $S^{th}$ cycle.
Thus, a station
receives a service for {\it L ms}
with a Goodput $G$, and then
waits
$(S-1) \cdot L\  ms$ before receiving TCP Data segments again.
In total the system provides a Goodput $G$ for all stations
during an interval of $S \cdot L \ ms$.

In scheduling strategy 3 (the MU one) where $S$ stations
transmit in a TXOP, every station has a Goodput $\frac{G}{S}$
during an interval of $L\  ms$. Overall the system provides
a Goodput $G$ to all the stations during every interval
of $L\ ms$.



\section{Goodput results}

\begin{figure}
\vskip 20cm
\includegraphics{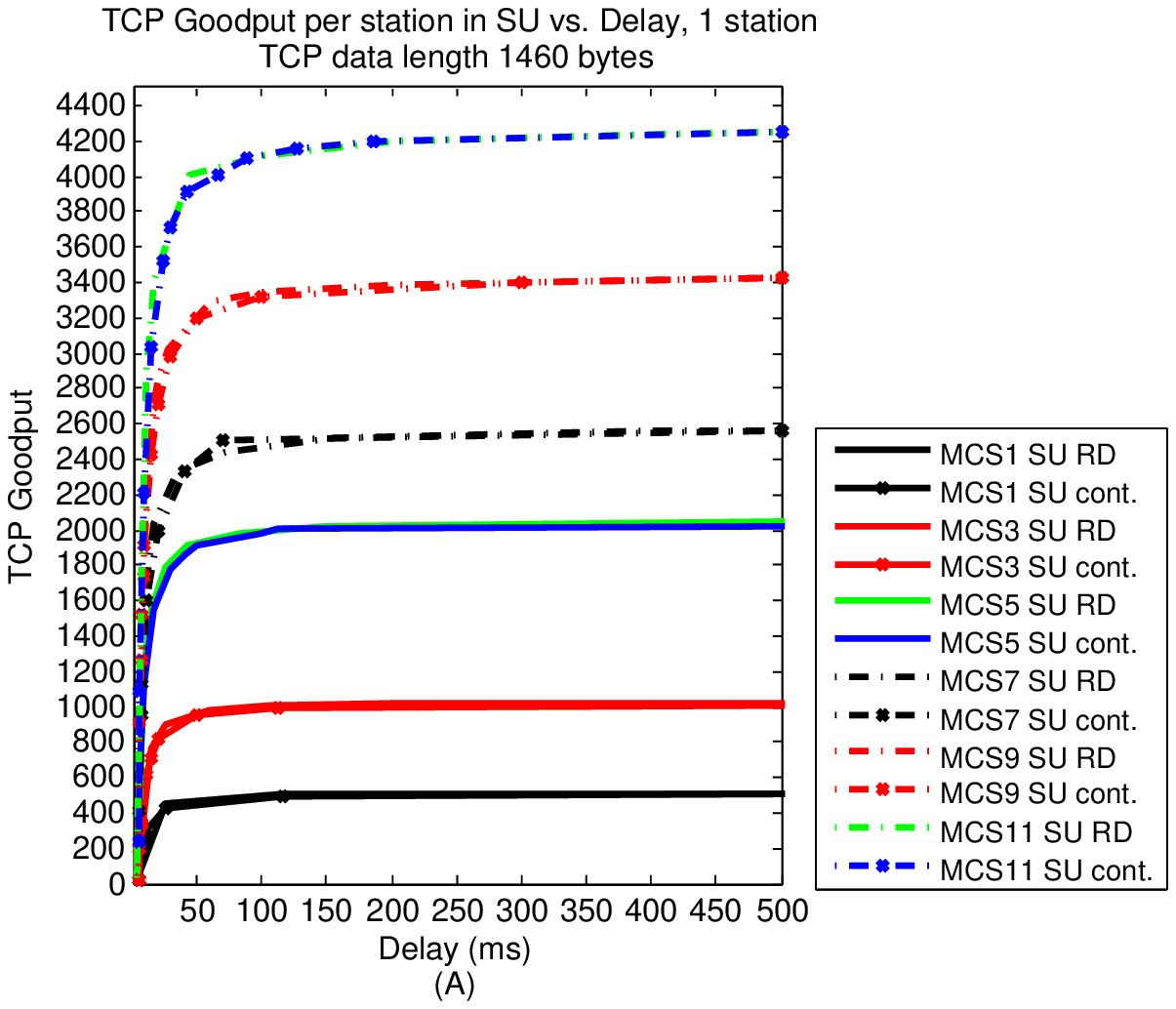}
\includegraphics{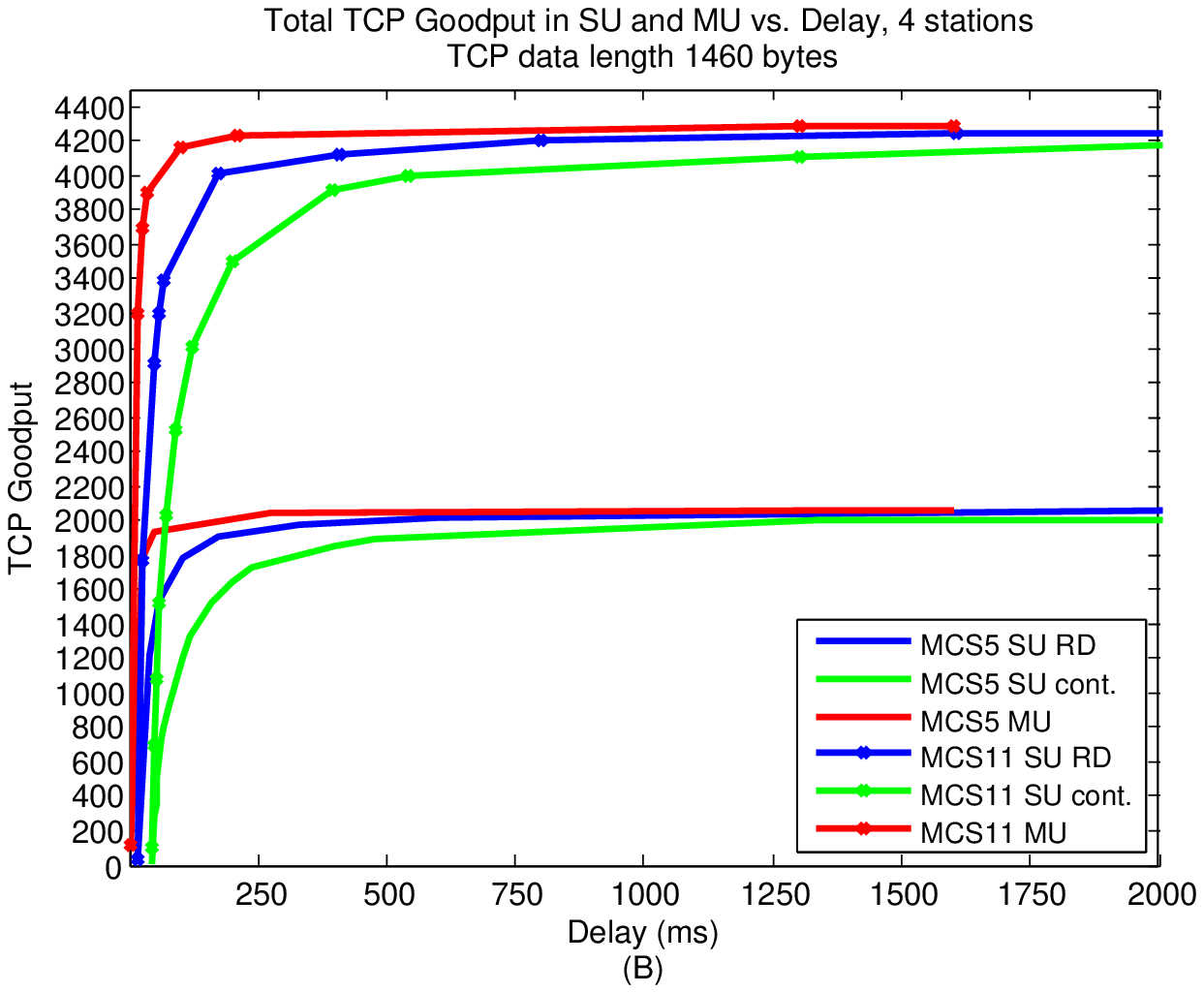}
\includegraphics{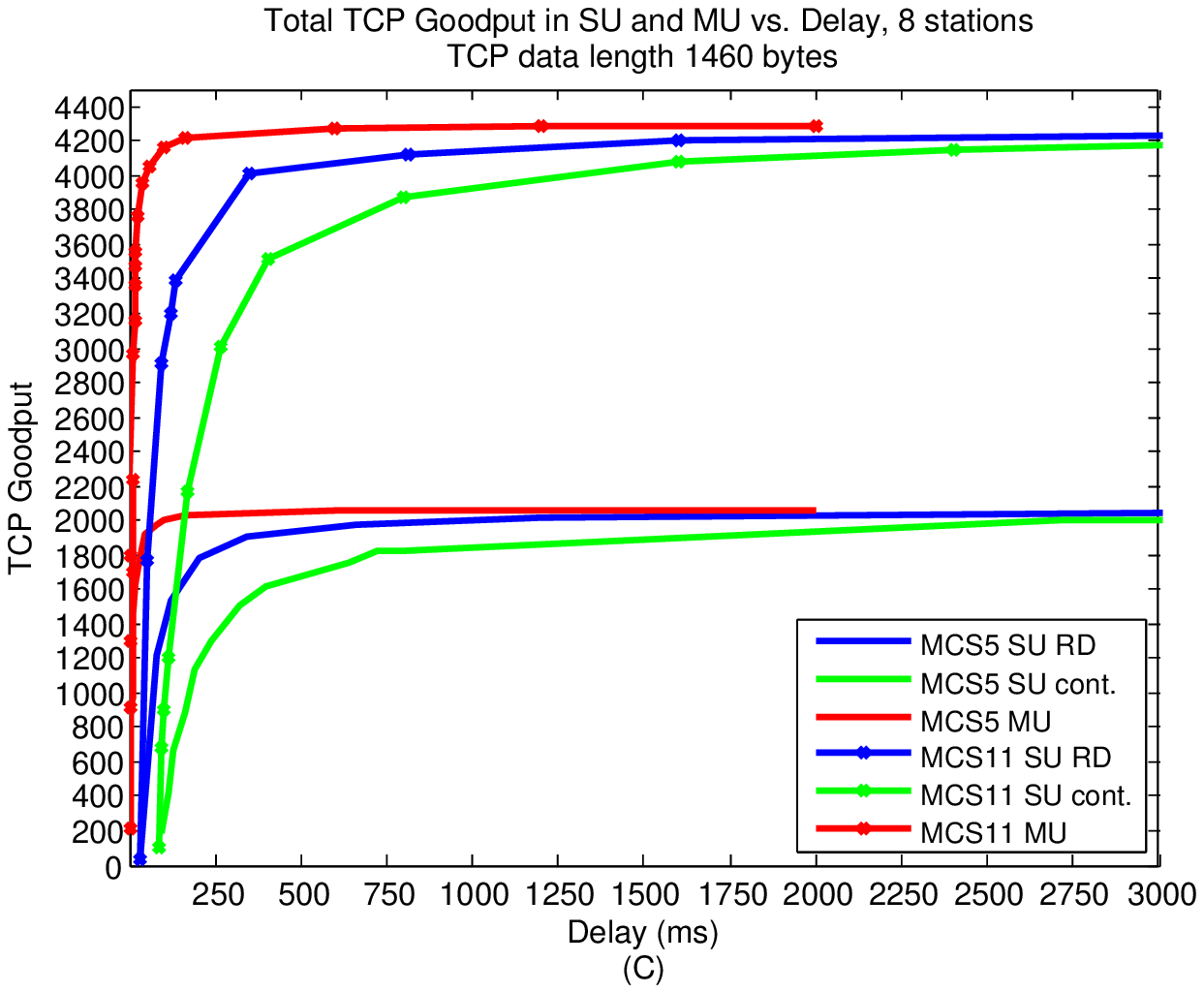}
\includegraphics{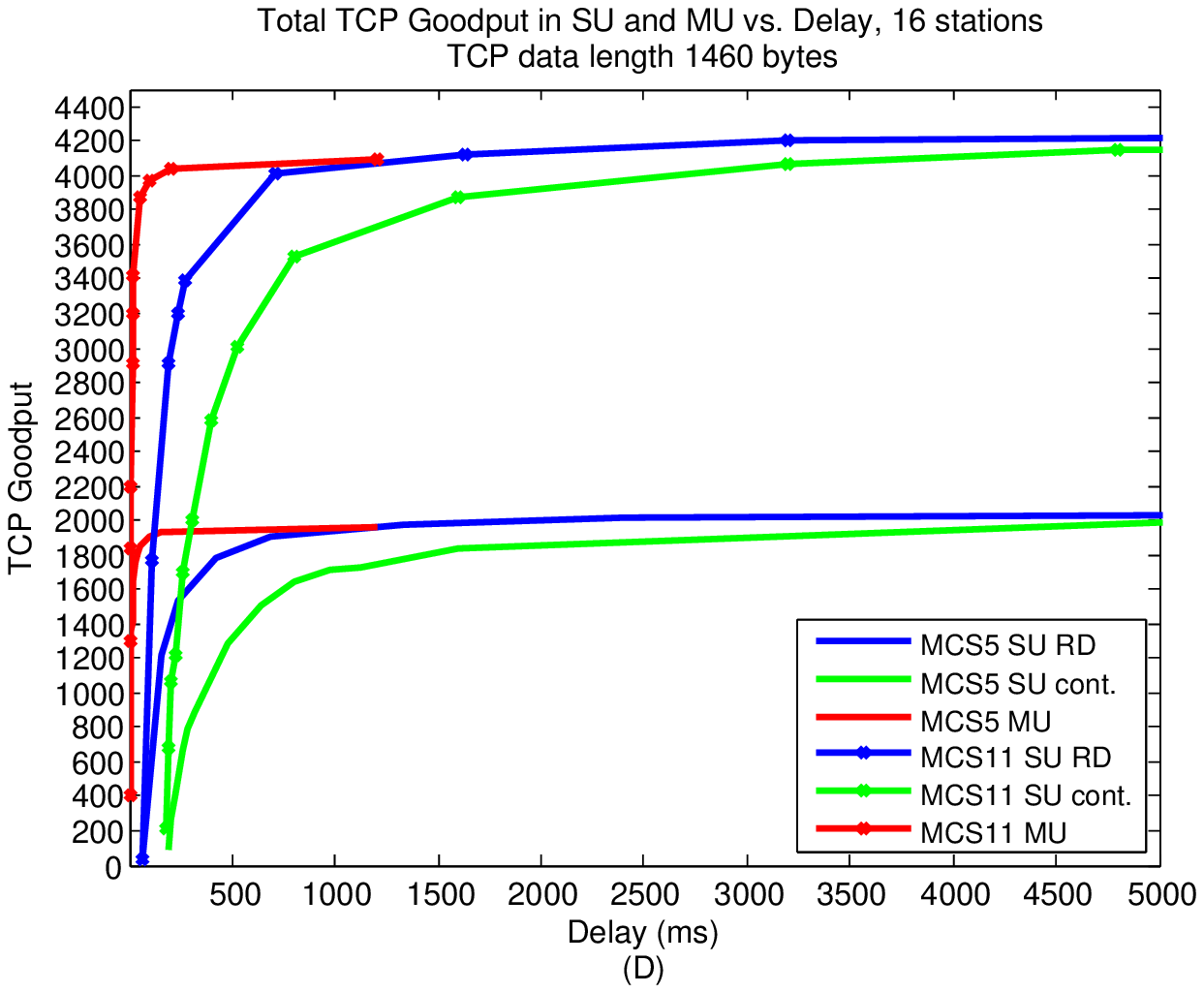}
\includegraphics{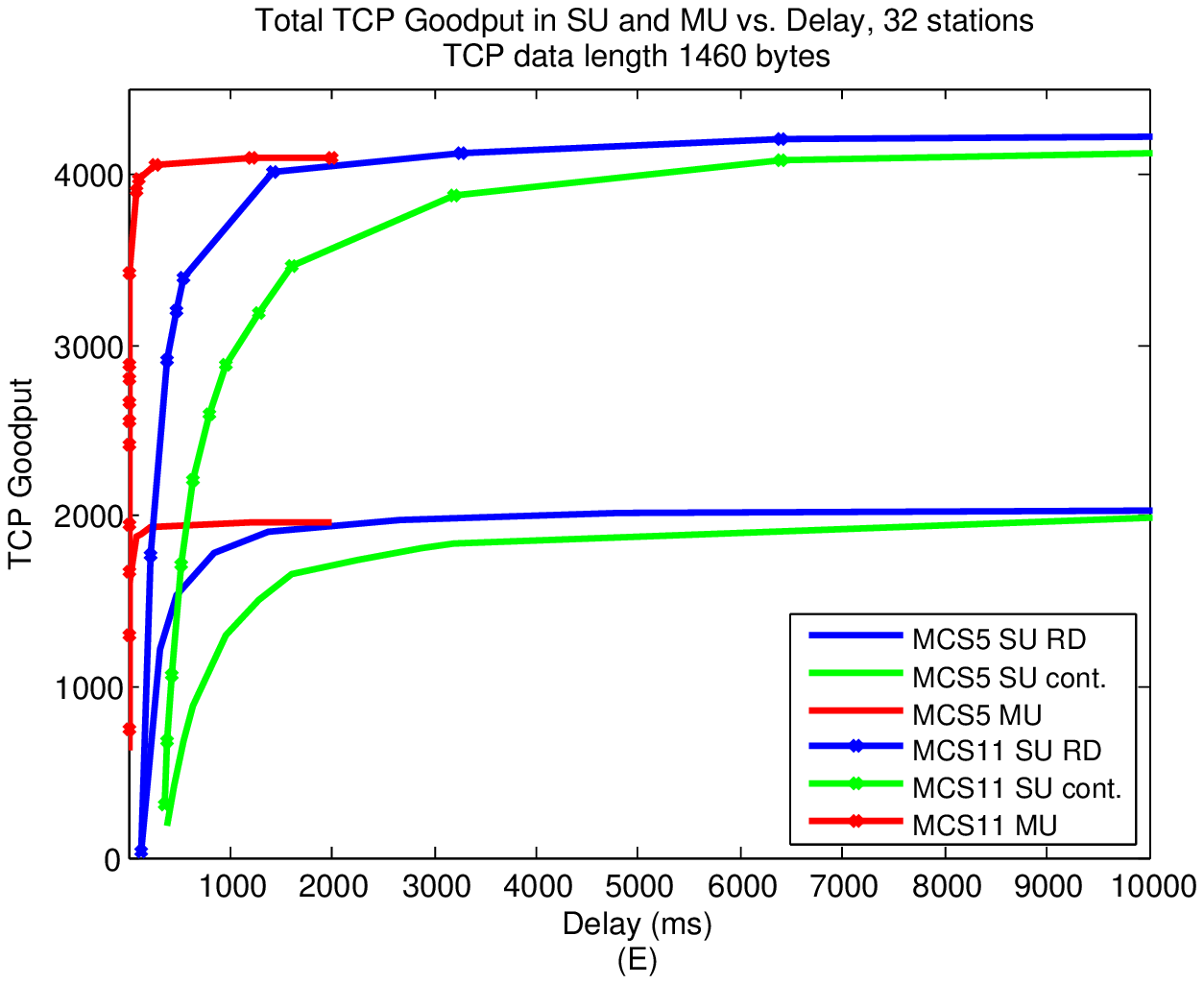}
\includegraphics{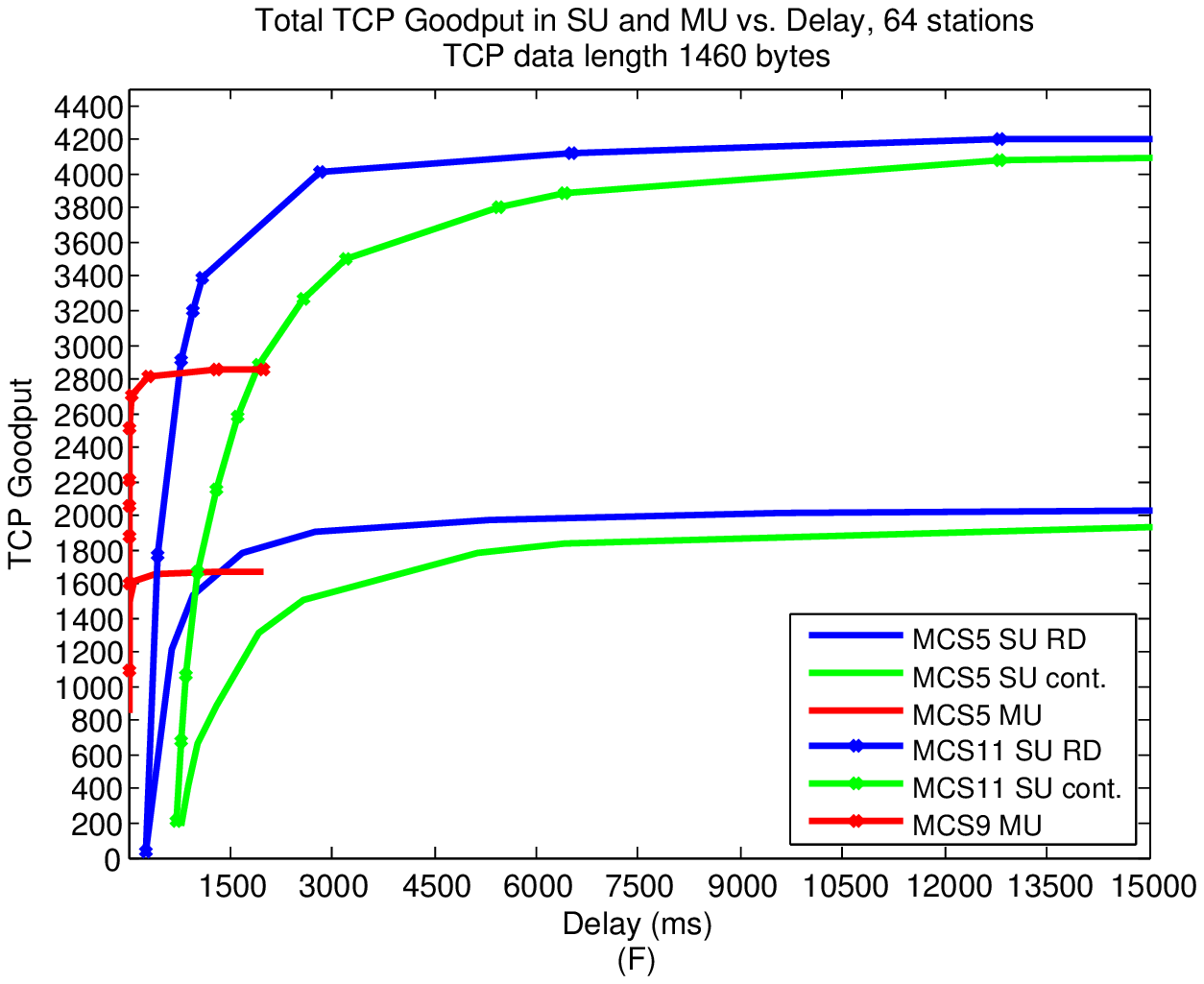}
\caption{Total TCP Goodput vs. delay in Scheduling strategies 1, 2 and 3 in MCS 5 and MCS 11. 1, 4, 8, 16, 32 and 64 stations. TCP data length 1460 bytes.} 
\label{fig:allsta}
\end{figure}

In Figure~\ref{fig:allsta} we plot 6 graphs showing the
Goodput
of the system vs. the delay (cycle length) for the cases of $S=1, 4, 8, 16, 32$
and 64 stations in Figures~\ref{fig:allsta}(A), (B), (C), (D), (E) and (F)
respectively, for TCP Data segments of length
1460 bytes and for the case where Delayed Acks
are not used, i.e. every TCP Ack acknowledges one
TCP Data segment. The results for TCP Data segments
of 464 and 208 bytes are similar.
As mentioned, the graphs for scheduling strategies 1 and 3
were obtained by analysis and simulation. The results for scheduling
strategy 2 were obtained by simulation only.

In Figure~\ref{fig:allsta}(A) we show results for a single
station and therefore only scheduling strategies 1 and 2 
are relevant. We show results for MCSs 1, 3, 5, 7, 9 and 11.
We see that the two scheduling strategies have similar
results because there are no collisions in scheduling strategy 2 - the
AP and the single station transmit alternately.

In Figure~\ref{fig:allsta}(B) we show results for 4 stations
and for MCSs 5 and 11, this time for all the scheduling strategies.
The results for all the other MCSs are similar.
We see that the maximum Goodput is received in MU for about $300 ms$
while in SU the same maximum Goodput is received in much longer
times, i.e. more TCP Data segments need to be transmitted.
Therefore, the MU strategy outperforms the SU strategies, while
using RD outperforms the contention based strategy.
We can therefore conclude that the MU uses the channel
more efficiently in this case, and enables a better performance
for TCP than SU.

The same result also holds for 8 stations, Figure~\ref{fig:allsta}(C).
In the case of 16 stations the MU strategy almost achieves
the maximum Goodput. The RD strategy achieves the maximum Goodput,
although in much larger delays. The MU strategy has
small PHY rates that do not enable transmission of many TCP Acks
due to the limit on transmission time of the HE UL MU A-MPDU frame
containing the TCP Acks. As a consequence the number of
TCP Data segments that can be transmitted in a TXOP
is relatively small. Therefore, it is not possible
to transmit as many TCP Data segments
in a TXOP as in the SU strategies, 
and the resulting Goodput is smaller.

Notice that the above phenomena is also observed in the case
of 32 stations, Figure~\ref{fig:allsta}(E). In Figure~\ref{fig:allsta}(F),
the case for 64 stations, the very small PHY rates in MU
cause the SU modes to outperform MU significantly.

\begin{figure}
\vskip 20cm
\includegraphics{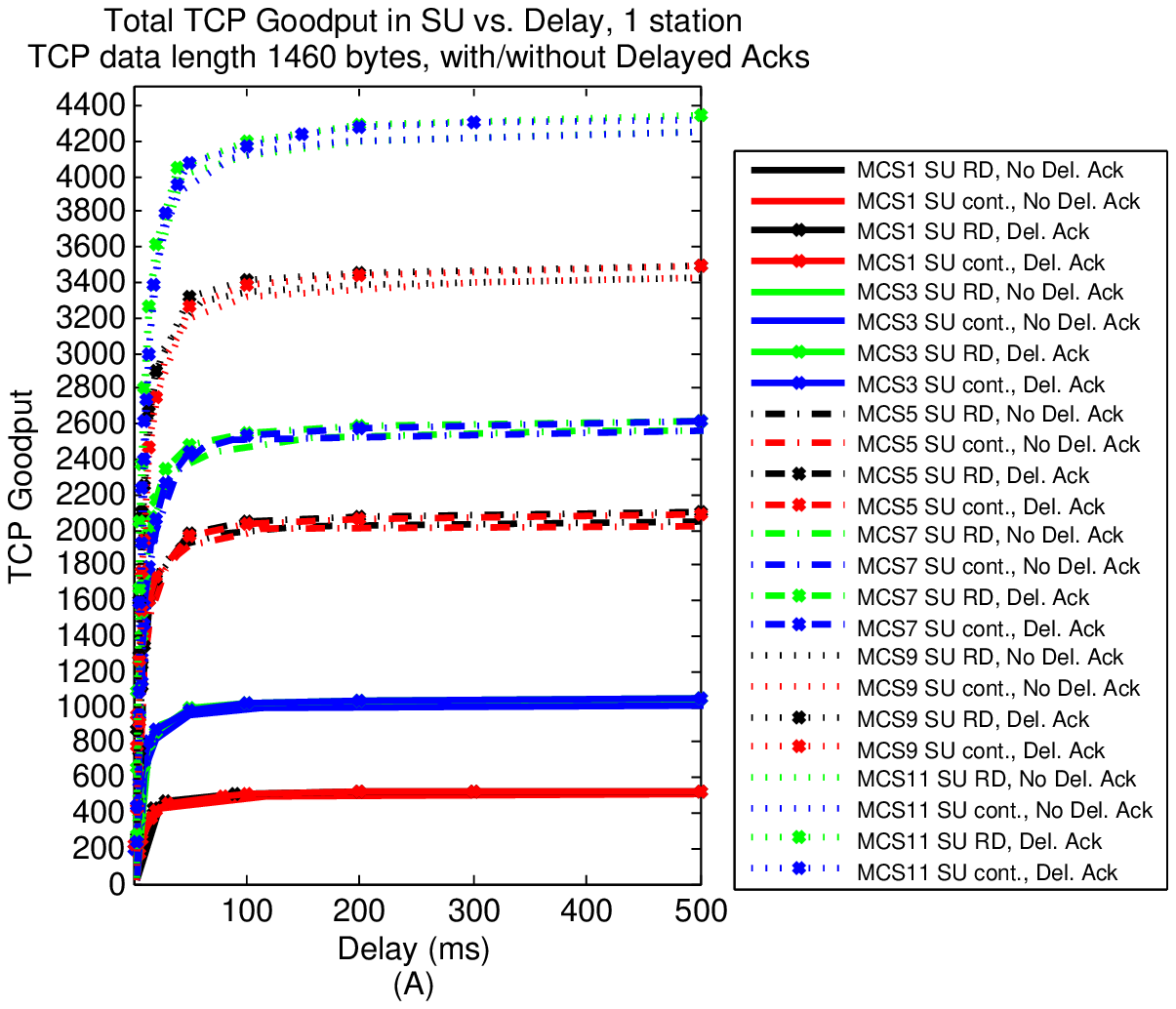}
\includegraphics{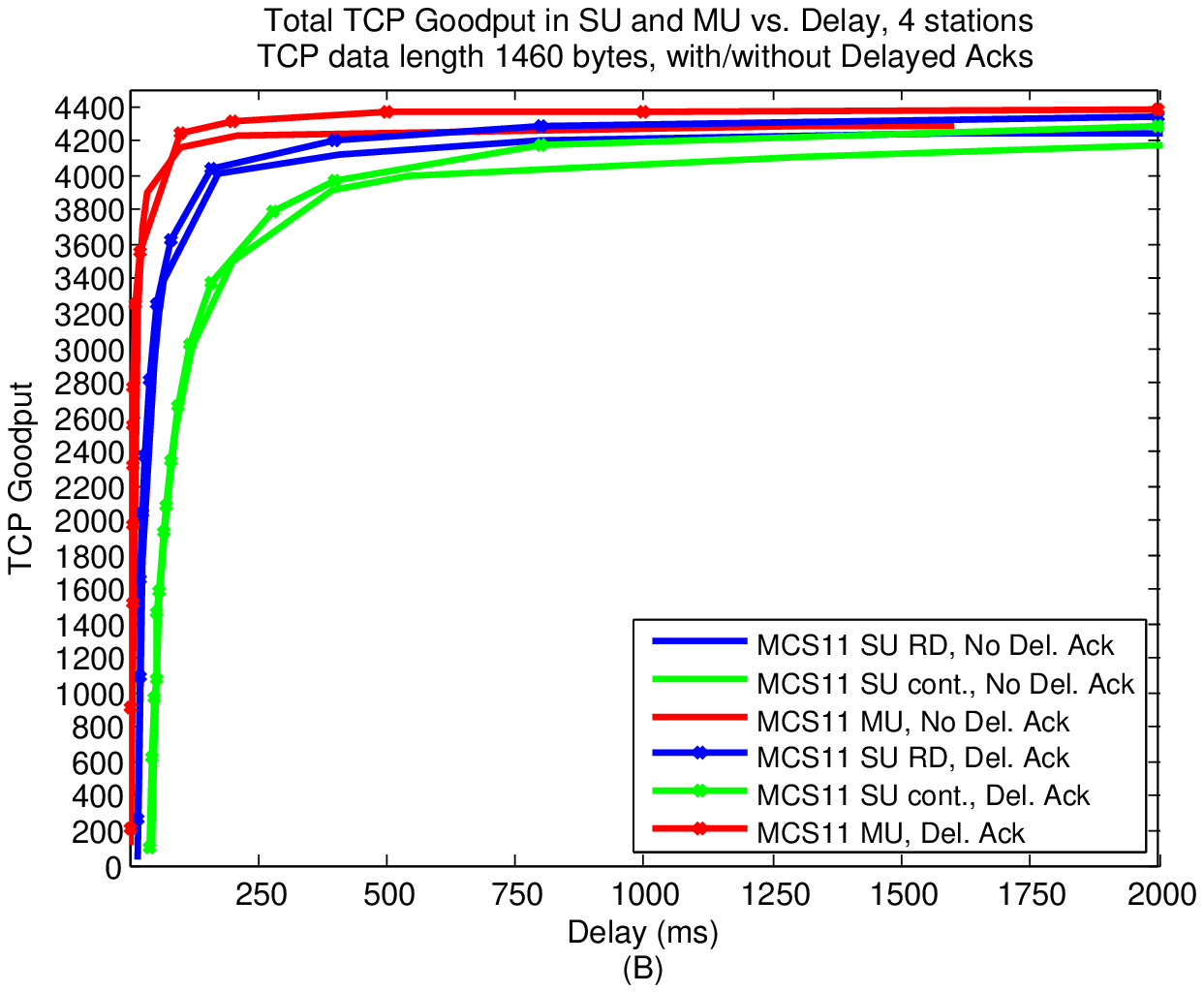}
\includegraphics{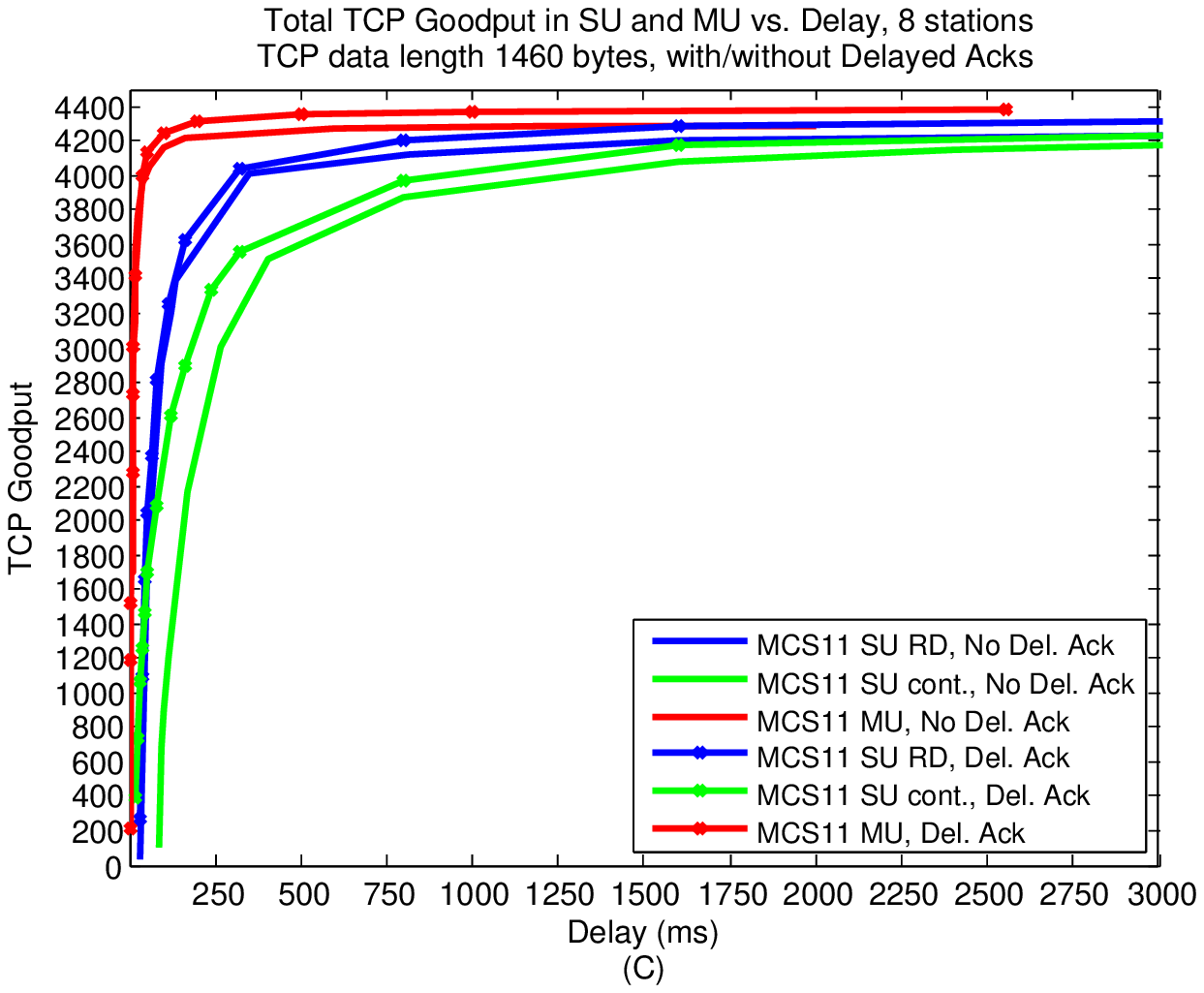}
\includegraphics{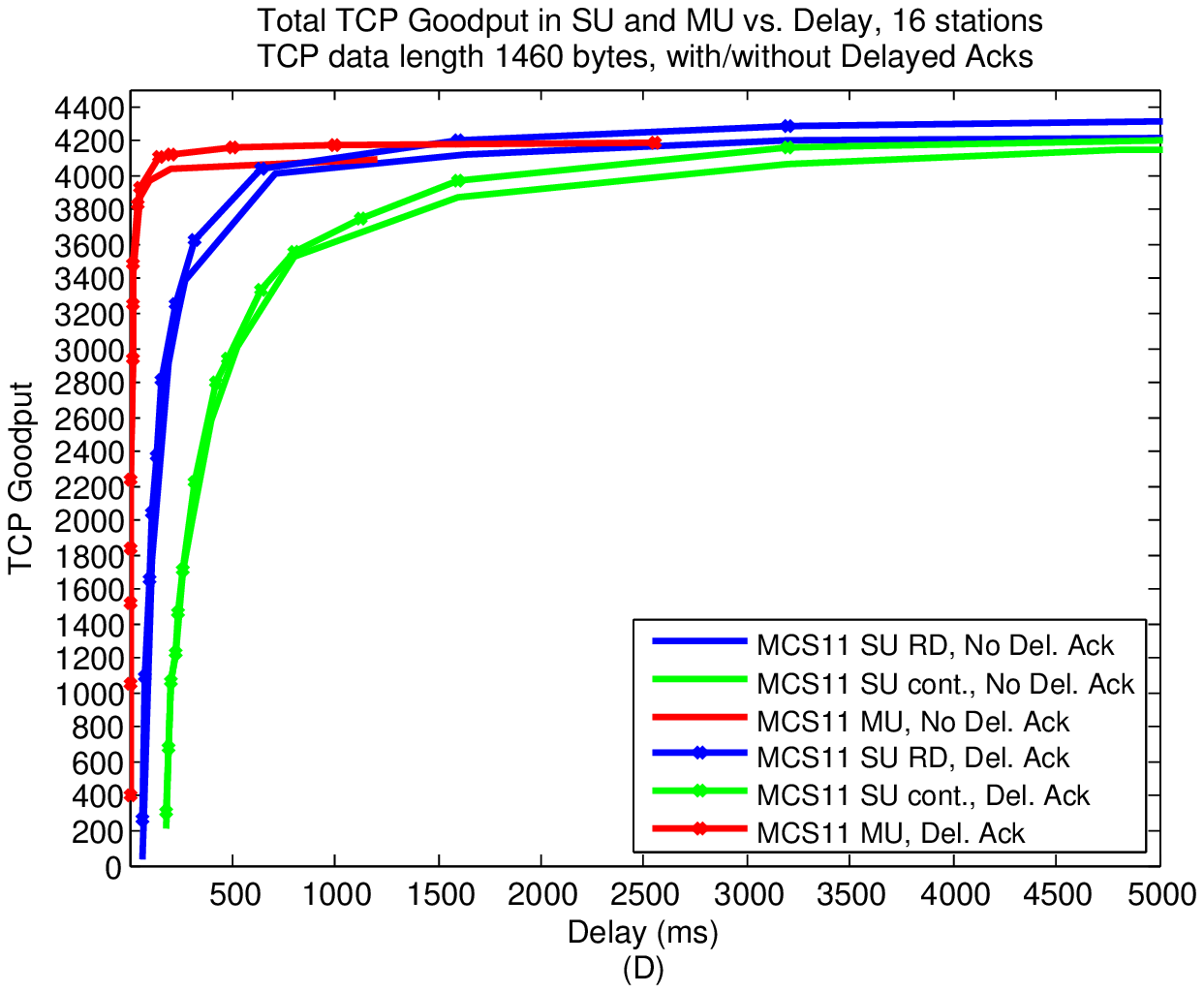}
\includegraphics{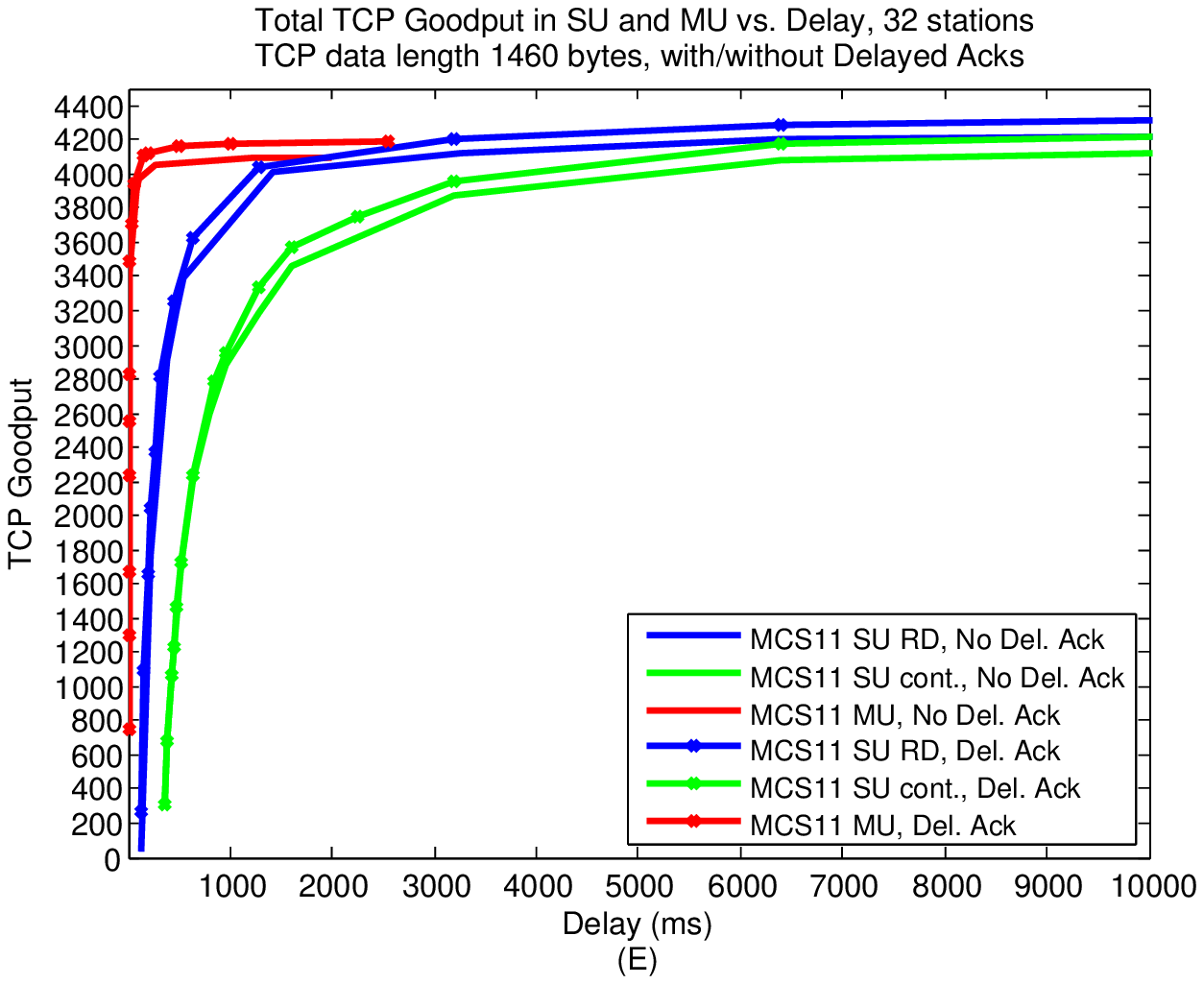}
\includegraphics{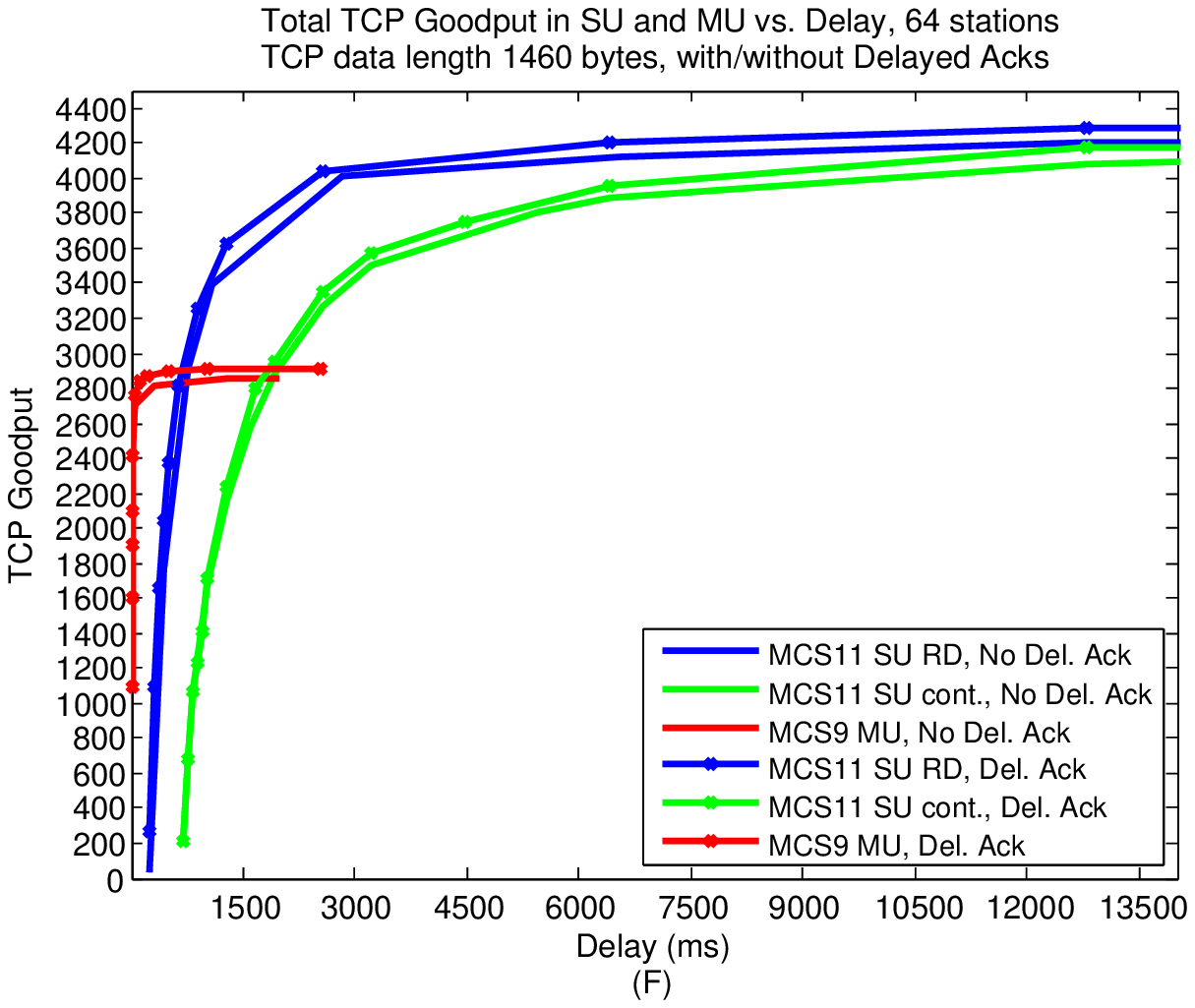}
\caption{Total TCP Goodput vs. delay in Scheduling strategies 1, 2 and 3 in MCS  11. 4, 8, 16, 32 and 64 stations. TCP data length 1460 bytes. Delayed Acks vs. No Delayed Ack.} 
\label{fig:delack}
\end{figure}

In Figure~\ref{fig:delack} we show the same results
as in Figure~\ref{fig:allsta}, this
time with results for the Delayed Acks feature.
For 4, 8, 16, 32 and 64 stations we show
results only for
MCS11. It can be concluded
that the improvement in the Goodput is only marginal.
Using Delayed Acks enables transmission of more TCP Data
segments in a cycle. However, there is no save in the
overhead of A-MPDUs and MPDUs containing TCP Data
segments. There is only a small save in the
overhead involved in the transmission of the TCP Acks,
which is marginal, and so is the Goodput gain.

\begin{figure}
\vskip 12cm
\includegraphics{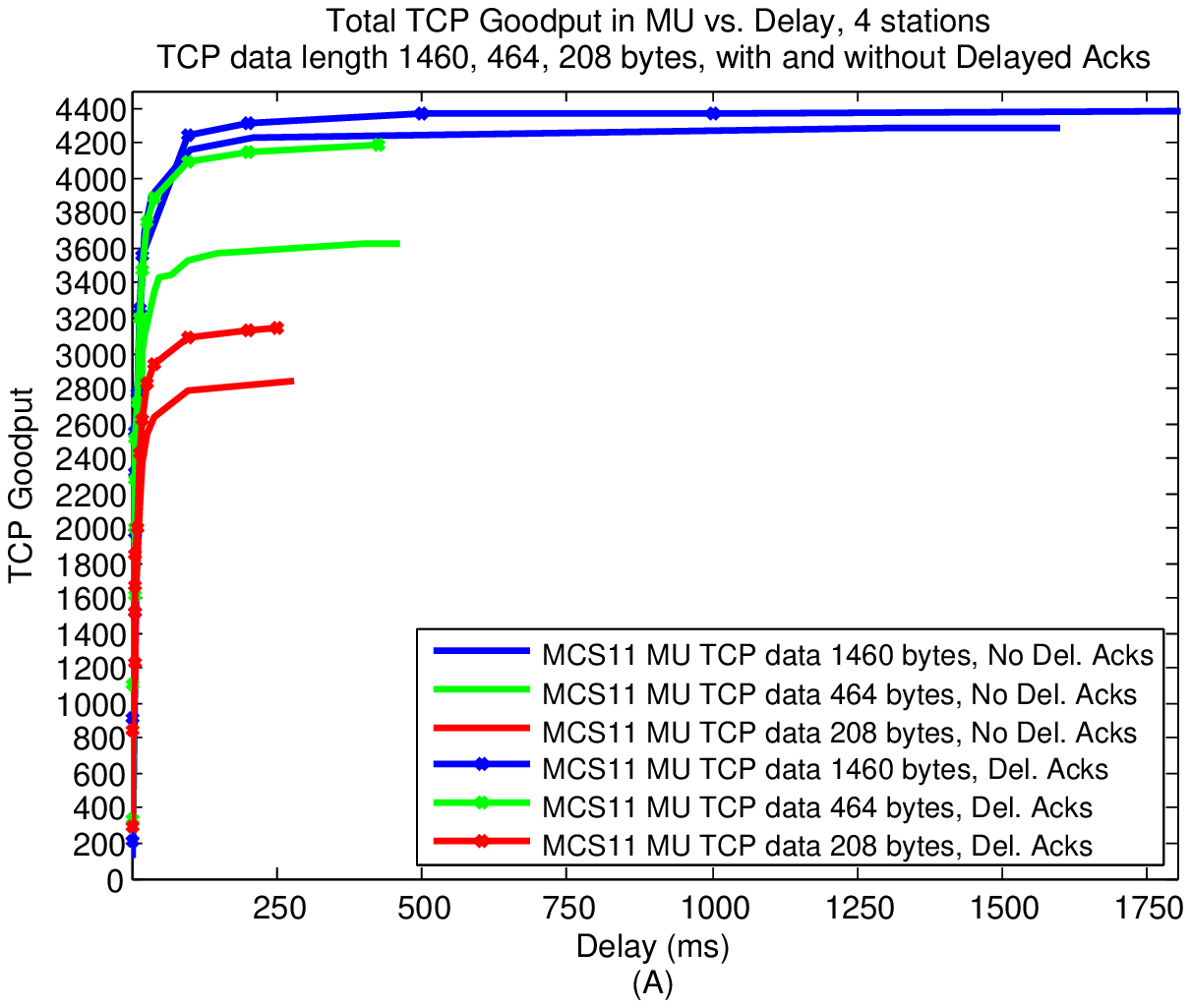}
\includegraphics{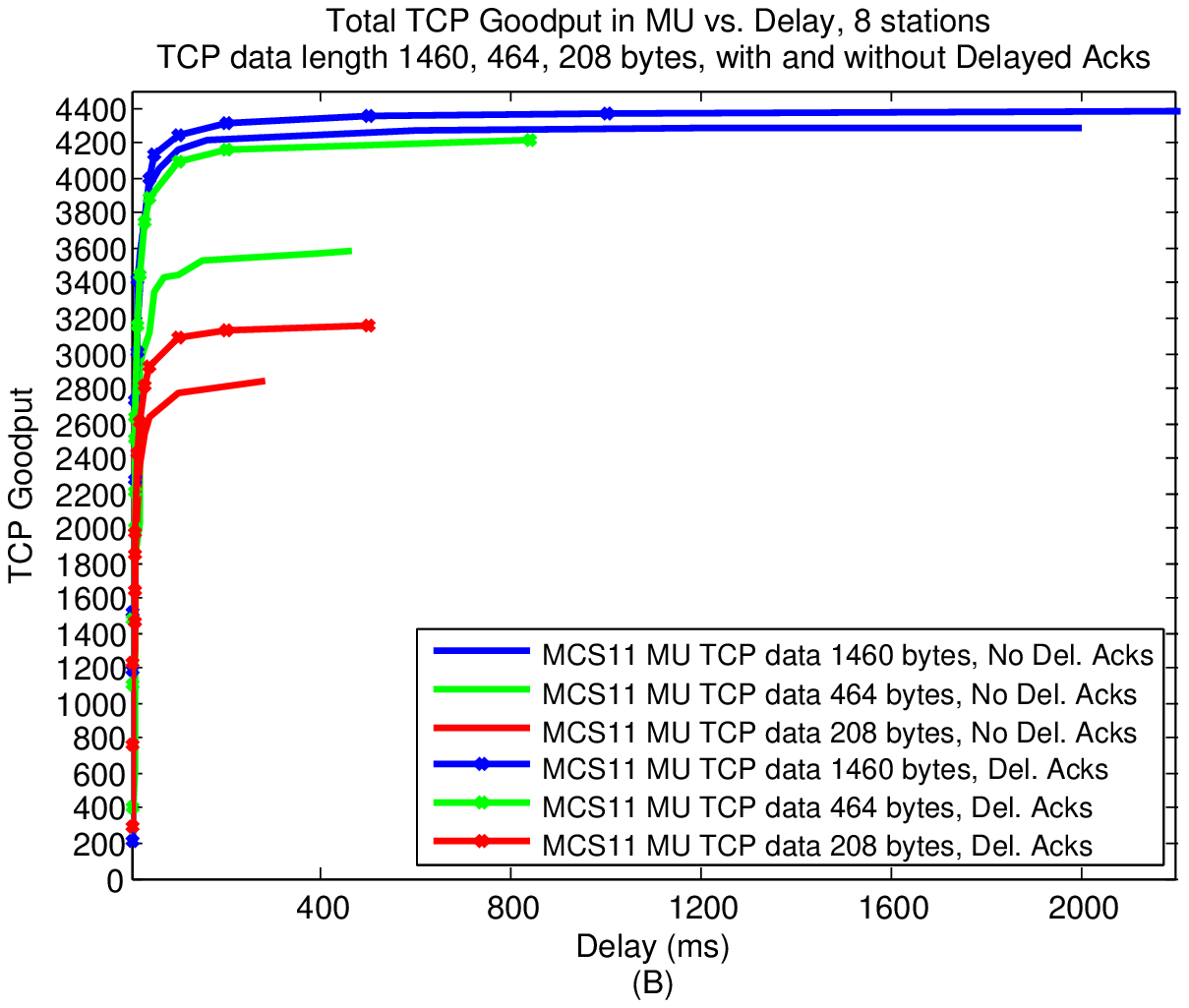}
\includegraphics{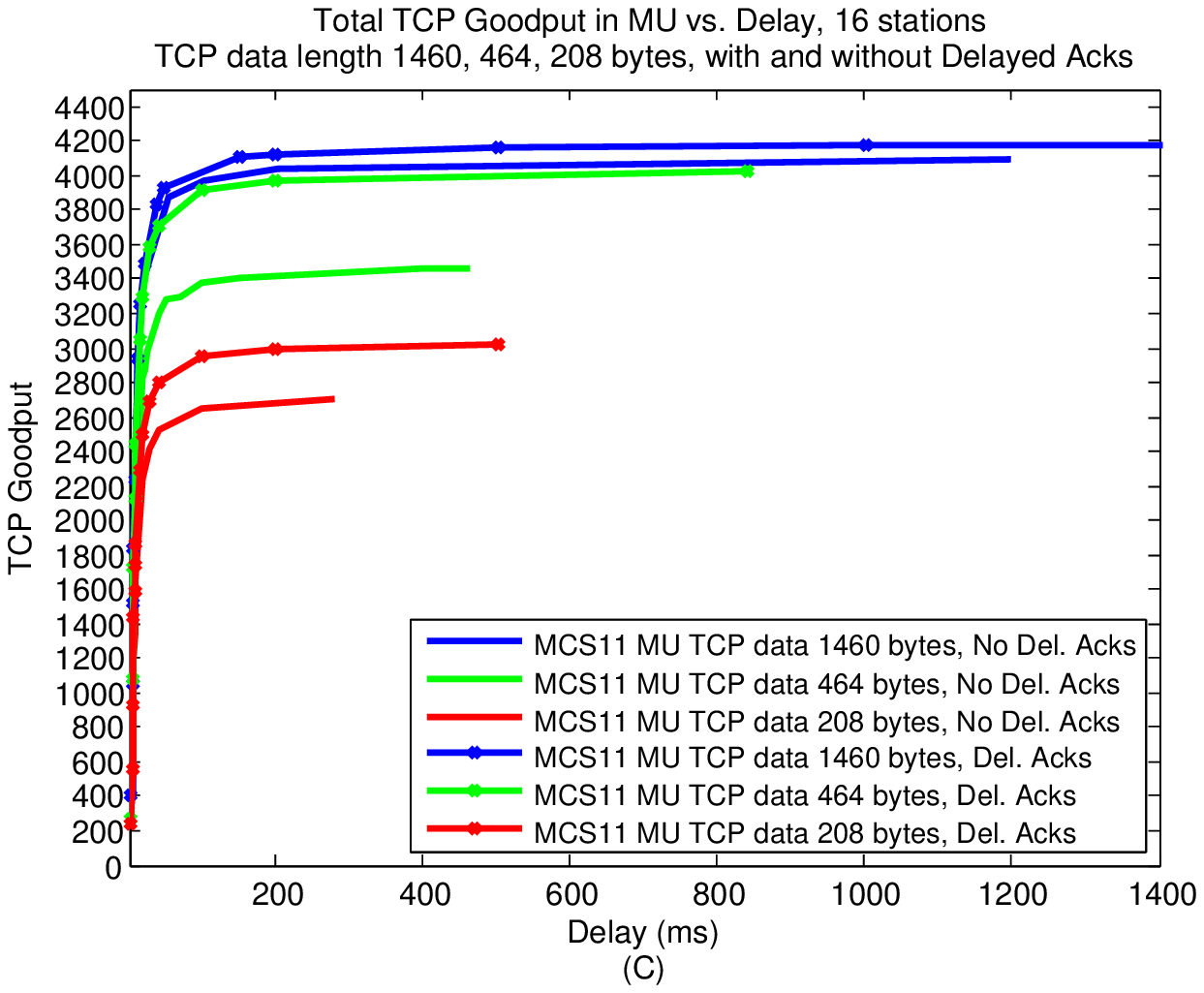}
\caption{Total TCP Goodput vs. delay in MU - scheduling strategy 3 . Number of stations: 4, 8 and 16. TCP data lengths 1460, 464 and 208 bytes.}
\label{fig:compmu}
\end{figure}

In Figure~\ref{fig:compmu} we show results for the
various TCP Data segments' sizes, 208, 464 and 1460 bytes
for MCS11, for the cases of 4, 8 and 16 stations
in Figures~\ref{fig:delack} (A), (B) and (C) respectively.
We also show results with and without Delayed Acks.
Since the number of TCP Acks that can be transmitted in a cycle
does not change, one can expect that as the length of the
TCP Data segments decreases, the length of the
respective cycles also decrease. This also is
true for the respective
Goodputs since the overhead of transmitting TCP
Ack segments remains unchanged. 

We see these expected results in Figure~\ref{fig:compmu}.
Notice that for all cases the curves end at the longest cycles
possible and these lengths decrease as the TCP Data segment
lengths decrease.

We can also see that while for TCP Data segments of 1460 bytes
the use of Delayed Acks results only in marginal Goodput
improvement, the other TCP Data segments' lengths
such as 464 and 208 bytes show significant
improvement. in the order of $15-20\%$. With short
TCP Data segments one can add many 
such segments without increasing the number
of MPDUs and A-MPDUs significantly, while greatly
increasing the number of TCP Data bytes transmitted. Therefore,
the ratio between the increase in the TCP data to the
increase in the A-MPDUs/MPDUs overheads is much better
than in the case of long TCP Data segments and the
increase in the Goodput is more significant.

\section{Summary}

In this paper we have introduced three scheduling strategies
for the transmission of TCP Data over the DL of an IEEE 802.11ax
system, where the AP is the TCP Data transmitter
and the stations are the receivers. Two of the strategies
are SU, and one strategy is an MU. We measured
the Goodput of the system as a function of the time
it takes the system to provide this Goodput.

We found that for up to 8 stations the MU strategy
outperforms those of SU, i.e. the maximum Goodput
is achieved in the MU strategy in much shorter time
intervals than in the SU strategies. For the case of
16 and 32 stations the MU strategy achieves almost the same
Goodput of the SU strategies, but does so in much
shorter time intervals. The SU strategies achieve a slightly
larger Goodput but with much longer time intervals. Therefore,
in these cases it is not clear which is the best strategy.
For the case of 64 stations the SU strategies are much better
than the MU because the latter has very small PHY rate 
channels.

Finally, we found that using Delayed Acks has only marginal
influence on the Goodput when transmitting long TCP Data
segments. The Delayed Acks feature results with significant
improvement in the achieved Goodput, in the order of 
when the TCP Data segments are short.

\section{Appendix}

In this appendix we show how to schedule $N$ TCP Data MSDUs
into MPDUs and A-MPDUs frames
in the HE DL TCP Data cycles of
various scheduling strategies,
in a way that minimizes the
MPDUs' overhead when the number of A-MPDUs is given. 
We first compute upper and lower limits
on the number of A-MPDUs needed for the transmission of
$N$ TCP Data MSDUs. For
every number of A-MPDUs in the range we show scheduling
that uses a minimal number of MPDUs, i.e. the smallest
MPDUs' overhead.

\noindent
We begin with several definitions.

\noindent
{\bf Definitions:}

\begin{enumerate}

\item
${\bf A_{Data}}$: The maximum possible number of TCP Data MSDUs in an MPDU.

\item
{\bf Full MPDU}: An MPDU containing $A_{Data}$ TCP Data MSDUs.

\item
{\bf Partial MPDU}: An MPDU containing less than $A_{Data}$ TCP Data MSDUs.

\item
${\bf F_{Data}}$: The maximum possible number of Full MPDUs in an A-MPDU.

\item
{\bf Full A-MPDU}: 
An A-MPDU containing the maximum possible
number of TCP Data MSDUs.

\item
{\bf Partial A-MPDU}: 
An A-MPDU that contains less than the maximum possible
number of TCP Data MSDUs.
 
\end{enumerate}

Consider an A-MPDU frame that contains $F_{DATA}$ Full MPDUs and
possibly one more Partial MPDU. The Partial MPDU
contains the maximum possible number
of MSDUs given the limit on transmission time
of a PPDU. We denote such a scheduling
of MSDUs within an A-MPDU by $F-construction$.

\noindent
{\bf Claim 1:} 
An A-MPDU with an F-construction is a Full A-MPDU.

\noindent
{\bf Proof:}
Let X and Y be the numbers of the MPDUs and MSDUs in
the F-construction respectively. Assume it is
possible to schedule $Y^{'}>Y$ MSDUs in an A-MPDU
given the limit on the transmission time of a PPDU.

It must hold that the $Y^{'}$ MSDUs are scheduled in the
A-MPDU within X or more MPDUs because it is immediately
seen that X is the minimum number of MPDUs necessary
for the transmission of $Y$ MSDUs.

It is not possible to schedule the $Y^{'}$ MSDUs in
X MPDUs only because this will violate the limit on the
transmission time of the PPDU containing the A-MPDU.
Therefore, at least X+1 MPDUs are required. This means
that the transmission time of the A-MPDU with $Y^{'}$
MSDUs is larger than that of an A-MPDU
with an F-construction by at least
the transmission time of one MPDUs' overhead and one MSDU.
However, this violates the given that it is not possible
to add even one more MSDU to an A-MPDU with
an F-construction without violating
the transmission time of a PPDU.

\qed

Let $X_{min} = \ceil{\frac{N}{A_{Data}}}$ be the minimum number of
MPDUs necessary to contain $N$ TCP Data MSDUs.
It is possible to schedule the MSDUs into
the MPDUs such that
$\ceil{\frac{N}{A_{Data}}}-1$ MPDUs are Full MPDUs and one more MPDU is
either a Full or a Partial MPDU.

Let $X_{upper} = \ceil{\frac{X_{min}}{F_{Data}}}$ be an upper limit
on the number of A-MPDUs that are 
needed to contain the $X_{min}$ MPDUs. 
$(X_{upper}-1)$ A-MPDUs contain $F_{Data}$ MPDUs and the
last A-MPDU possibility contains less than $F_{Data}$ MPDUs
and possibility one Partial MPDU.

Let $X_{lower}$ be a lower limit
on the number of A-MPDUs needed to transmit the
$N$ TCP Data MSDUs. One possibility for scheduling the $N$
TCP Data MSDUs in these $X_{lower}$ A-MPDUs is by 
defining $(X_{lower}-1)$
Full A-MPDUs and possibly one Partial A-MPDU.

Notice that by using $X_{upper}$ A-MPDUs one uses the
smallest amount of overhead caused by MPDUs and the
largest amount of overhead caused by A-MPDUs.
The MPDU's overhead is the
MAC Header, MPDU Delimiter and the FCS fields. The 
A-MPDU overhead for scheduling strategies  1 and 2
is $Pr(6(A))+Pr(6(B))+T(BAck)
+2 \cdot SIFS$. For scheduling strategy 3 the preambles are
$Pr(6(C))$ and $Pr(6(D))$.

By using $X_{lower}$ A-MPDUs one uses the largest amount
of overhead caused by MPDUs and the smallest amount of
overhead caused by A-MPDUs.
To find the maximum Goodput when transmitting
$N$ TCP Data MPDUs, one needs to review
all numbers of A-MPDUs $X$,
$X_{lower} \le X \le X_{upper}$, and determine
the minimum MPDUs' overhead when using $X$ A-MPDUs.
We then need to find 
the minimum sum
of overheads of both A-MPDUs and MPDUs
for all $Xs$ in the range.

In the following we show 
the scheduling that results with the smallest amount
of MPDUs' overhead 
given $N$
TCP Data MSDUs and $X$ A-MPDUs,

For this purpose we
now define the following scheduling $\alpha$ of $N$ TCP Data MSDUs
into $X$ A-MPDUs, $X_{lower} \le X \le X_{upper}$.

\noindent
{\bf Scheduling $\alpha$:}

\begin{itemize}

\item
$X=X_{upper}:$  $(X_{upper}-1)$ 
A-MPDUs contain only $F_{Data}$ Full MPDUs.
The last A-MPDU contains as many as possible Full MPDUs
and possibly one Partial MPDU.

\item
$X_{lower} \le X \le (X_{upper}-1)$ : All the A-MPDUs contain
$F_{DATA}$ Full MPDUs.
The remaining MSDUs are used to construct as many as possible
Full A-MPDUs (F-construct).

\end{itemize}

\noindent
{\bf Claim 2:} 
Assume $X$ A-MPDUs,
$X_{lower} \le X \le (X_{upper}-1)$ in scheduling $\alpha$,
such that every A-MPDU contains $F_{DATA}$ Full MPDUs.
Scheduling $\alpha$ then contains
the minimal number of Partial MPDUs
needed for the scheduling of the remaining
$M=N-A_{data} \cdot F_{Data} \cdot X$ TCP Data
MSDUs in the $X$ A-MPDUs.

\noindent
{\bf Proof:}
Assume that every Partial MPDU in an A-MPDU
with $F_{Data}$ Full MPDUs can contain
at most $M^{'}$ MSDUs. To schedule $M$ MSDUs in such Partial
MPDUs one needs at least $\ceil{\frac{M}{M^{'}}}$ such
MPDUs, which is the number of MPDUs that scheduling $\alpha$ uses.

\qed

\noindent
{\bf Claim 3:}
Let $P$ be the number of MPDUs defined by scheduling $\alpha$
when scheduling $N$ TCP Data MSDUs in X A-MPDUs, 
$X_{lower} \le X \le (X_{upper}-1$). $P$ is then the minimal
number of MPDUs necessary to schedule $N$ TCP Data MSDUs in the $X$ A-MPDUs.

\noindent
{\bf Proof:}
Assume
that $N= A_{DATA} \cdot F_{DATA} \cdot X + M, M \ge 0$
and thus $M$ is the number of MSDUs in Partial MPDUs in scheduling
$\alpha$.

Assume on the contrary
that there is another scheduling $\beta$ in which the
$N$ TCP Data MSDUs are scheduled in $X$ A-MPDUs within less than $P$ MPDUs.

Notice that if an A-MPDU in scheduling $\beta$ contains
two or more Partial MPDUs then it is possible to re-arrange the
scheduling of the MSDUs within the A-MPDU such that
the number of MPDUs is not changed and that all the MPDUs
in the A-MPDU are Full MPDUs except possibly one Partial MPDU.

We re-arrange all the A-MPDUs in 
scheduling $\beta$ as described above. 
We go then through the A-MPDUs
and for every A-MPDU number $t$, $1 \le t \le X$,
we do the following: If A-MPDU $t$ does not contain $F_{DATA}$
Full MPDUs, we borrow MSDUs from Partial MPDUs in other
A-MPDUs and schedule them in A-MPDU $t$ such that we form
$F_{DATA}$ Full MPDUs in A-MPDU $t$. Notice that such a process
does not increase the overall number of MPDUs;
if a new MPDU is generated in A-MPDU $t$, a Partial MPDU
in another A-MPDU must be canceled. 

The above process can be performed over all the A-MPDUs
as there are at least $A_{DATA} \cdot F_{DATA} \cdot X$
MSDUs to schedule. During the process the number of MPDUs
does not increase, but does not decrease due
to the optimality of scheduling $\beta$. By Claim 2
scheduling $\beta$ cannot have a smaller number of MPDUs than
scheduling $\alpha$.

\qed

\noindent
{\bf Lemma 1:}
For $X$ A-MPDUs, $X_{lower} \le X \le X_{upper}$ and $N$ TCP
Data MSDUs, scheduling $\alpha$ results in the smallest
MPDUs' overhead.

\noindent
{\bf Proof:}
For every $X$ in the given range, scheduling $\alpha$ generates
the smallest possible number of MPDUs needed for the
scheduling of $N$ TCP Data MSDUs. Therefore, scheduling $\alpha$
results in the smallest MPDUs' overhead.

\qed

\clearpage


\bibliographystyle{abbrv}
\bibliography{main}

\begin{thebibliography}{10}

\bibitem{IEEEBase1}
\newblock{IEEE Std. 802.11$^{TM}$-2016},
\newblock{IEEE Standard for Information Technology - 
Telecommunications and Information Exchange between Systems - Local
and Metropolitan Area Networks - Specific Requirements. Part 11:
Wireless LAN Medium Access Control (MAC) and Physical Layer (PHY)
Specifications},
\newblock{IEEE, NewYork, (December 2016)}


\bibitem{IEEEax}
\newblock{IEEE P802.11ax$^{TM}$/D1.4},
\newblock{IEEE Draft Standard for Information Technology - 
Telecommunications and Information Exchange between Systems - Local
and Metropolitan Area Networks - Specific Requirements. Part 11:
Wireless LAN Medium Access Control (MAC) and Physical Layer (PHY)
Specific requirements. }
\newblock{IEEE, NewYork, (2017)}

\bibitem{IEEEac}
\newblock{IEEE Std. 802.11ac$^{TM}$-2013},
\newblock{IEEE Standard for Information Technology - 
Telecommunications and Information Exchange between Systems - Local
and Metropolitan Area Networks - Specific Requirements. Part 11:
Wireless LAN Medium Access Control (MAC) and Physical Layer (PHY)
Specific requirements. Amendment 4: Enhancements for Very
High Throughput for Operation in Bands below 6 GHz},
\newblock{IEEE, NewYork, (2013)}

\bibitem{PS}
E. Perahia, R. Stacey,
\newblock{Next Generation Wireless LANs: 802.11n and 802.11ac,}
\newblock{2nd Edition, Cambridge Press, 2013 }

\bibitem{KKL}
E. Khorov, A. Kiryanov, A. Lyakhov,
\newblock{IEEE 802.11ax: How to Build High Efficiency WLANs,}
\newblock{Int. Conf. on Eng. and Telecommunication (2015) 14-19}

\bibitem{AVA}
M. S. Afaqui, E. G. Villegas, E. L. Aguilera,
\newblock{IEEE 802.11ax: Challenges and Requirements for Future
High Efficiency WiFi,}
\newblock{IEEE Wireless Communications 99 (2016) 2-9}

\bibitem{DCC}
D. J. Deng, K. C. Chen, R. S. Cheng,
\newblock{IEEE 802.11ax: Next Generation Wireless Local Area Networks,}
\newblock{10th Int. Conf. on Heterogeneous Networking for Quality, Security and Robustness (QSHINE), (2014) 77-82}

\bibitem{B}
B. Bellalta, 
\newblock{IEEE 802.11ax: High-efficiency WLANs,}
\newblock{IEEE Wireless Communications, 23(1) (2016) 38-46}

\bibitem{SA2}
O. Sharon, Y. Alpert,
\newblock{Scheduling strategies and Throughput optimization for
the Uplink for IEEE 802.11ax and IEEE 802.11ac based networks,}
\newblock{Wireless Sensor Networks,9 (2017) pp. 250-273}

\bibitem{SA3}
O. Sharon, Y. Alpert,
\newblock{Scheduling strategies and throughput optimization for
the Downlink for IEEE 802.11ax and IEEE 802.11ac based networks,}
\newblock{Submitted, Physical Communications}

\bibitem{SA1}
O. Sharon, Y. Alpert,
\newblock{Single User MAC level Throughput comparision: IEEE 802.11ax vs. IEEE
802.11ac,}
\newblock{Wireless Sensor Networks, 9 (2017), pp. 166-177}


\bibitem{KCC}
R. Karmakar, S. Chattopadhyay, S. Chakraborty,
\newblock{Impact of IEEE 802.11n/ac PHY/MAC High Throughput
Enhancement over Transport/Application layer protocols - A Survey,}
\newblock{IEEE Communication surveys and tutorials (2017)}

\bibitem{MKA}
Miorandi, D., Kherani, A. A. and Altman, E. (2006)
\newblock{ A Queueing model for HTTP traffic over IEEE 802.11 WLANs.}
\newblock{Computer Networks, 50, 63-79.}


\bibitem{BCG1}
Bruno, R., Conti, M. and Gregori, E. (2005)
\newblock{Throughput Analysis of UDP and TCP Flows in IEEE 802.11b WLANs: A Simple Model and its Validation.}
\newblock{Workshop on Techniques, Methodologies and Tools for Performance Evaluation of Complex Systems, 2005, 54-63.}


\bibitem{BCG2}
Bruno, R., Conti, M. and Gregori, E. (2008)
\newblock{Throughput Analysis and Measurements in IEEE 802.11 WLANs with
TCP and UDP Traffic Flows.}
\newblock{IEEE Trans. on Mobile Computing, 7, 171-186.}


\bibitem{KAMG}
Kumar, A., Altman, E., Miorandi, D. and Goyal, M. (2007)
\newblock{New Insights from a Fixed Point Analysis of Single Cell IEEE 802.11 WLANs.}
\newblock{IEEE/ACM Trans. on Networking, 15, 588-601.}


\bibitem{QLYY}
Q. Qu, B. Li, M. Yang, Z. Yan, 
\newblock{An OFDMA based Concurrent Multiuser MAC for Upcoming IEEE 802.11ax,}
\newblock{IEEE Wireless Comm. and Networking Conf. Workshops (WCNCW) (2015) 136-141}


\bibitem{LLYQYZY}
W. Lin, B. Li, M. Yang, Q. Qn, Z. Yan, X. Zuo, B. Yang,
\newblock{Integrated Link-System level Simulation Platform for the
Next Generation WLAN - IEEE 802.11ax,}
\newblock{IEEE Globecom (2016)}


\bibitem{LDC}
J. Lee, D. J. Deng, K. C. Chen,
\newblock{OFDMA-based hybrid channel access for IEEE 802.11ax WLAN,}
\newblock{unpublished.}

\bibitem{KBPSL}
M. Karaca, S. Bastani, B. E. Priyanto, M. Safavi, B. Landfeldt,
\newblock{Resource Management for OFDMA based Next Generation 802.11ax
WLANs,}
\newblock{9th IFIP Wireless and Mobile Networking Conf. (WMNC) (2016)}


\bibitem{JS}
V. Jones, H. Sampath,
\newblock{Emerging technologies for WLAN,}
\newblock{IEEE Commu. Mag., 5 (2015) 141-9 }

\bibitem{RFBBO}
L. Sanabria-Russo, A. Faridi, B. Bellalta, J. Barcelo, M. Oliver,
\newblock{Future evolution of CSMA protocols for the IEEE 802.11
standard}
\newblock{IEEE Int. Conf. on Comm. (ICC) (2013) 1274-9 }

\bibitem{RBFB}
L. Sanabria-Russo, J. Barcelo, A. Faridi, B. Bellalta,
\newblock{WLANs throughput improvement with CSMA/ECA,}
\newblock{IEEE Conf. on Computer Comm. Workshops (INFOCOM WKSHPS)
(2014) 125-6 }

\bibitem{HYSG}
Y. He, R. Yuan, J. Sun, W. Gong,
\newblock{Semi-Random Backoff: Towards resource reservation for
channel access in wireless LANs,}
\newblock{IEEE Int. Conf. on Network Protocols (ICNP) (2009) 21-30 }

\bibitem{KLL}
E. Khorov, V. Loginov, A. Lyakhov,
\newblock{Several EDCA Parameters Sets for Improving Channel Access in
IEEE 802.11ax Networks,}
\newblock{Int. Symposium on Wireless Communication Systems (ISWCS) (2016) 419-423}


\bibitem{SA10}
O. Sharon, Y. Alpert,
\newblock{Coupled IEEE 802.11ac and TCP performance evaluation in various aggregation schemes and Access Categories,}
\newblock{Computer Networks 100 (2016) 141-156}

\bibitem{SA11}
O. Sharon, Y. Alpert,
\newblock{Couples IEEE 802.11ac and TCP Goodput improvement using Aggregation and Reverse Direction,}
\newblock{Wireless Sensor networks 8(7) (2016) 107-136}

\bibitem{SA12}
O. Sharon, Y. Alpert,
\newblock{Comparison between TCP scheduling strategies in IEEE 802.11ac based Wireless networks,}
\newblock{Ad Hoc Networks 61C (2017) pp. 95-113}

\bibitem{SA}
O. Sharon, Y. Alpert,
\newblock{MAC level Throughput comparison: 802.11ac vs. 802.11n,}
\newblock{Physical Communication Journal 12 (2014) 33-49 }

\bibitem{IEEEn}
\newblock{IEEE Std. 802.11$^{TM}$-2012},
\newblock{Standard for Information Technology - 
Telecommunications and Information Exchange between Systems - Local
and Metropolitan Area Networks - Specific Requirements. Part 11:
Wireless LAN Medium Access Control (MAC) and Physical Layer (PHY)
specifications,}
\newblock{IEEE, NewYork, (2012)}

\bibitem{ns2}
\newblock{Network Simulator 2 (NS-2). Available:}
\newblock{http://www.isi.edu/nsnam/ns.}

\bibitem{M}
E. Modiano,
\newblock{An adaptive alogorithm for optimizing the packet size used in wireless ARQ protocols,}
\newblock{Wirel. Netw. 5 (1999) 279-286.}

\bibitem{YZZZ}
F. Yang, Q. Zhang, W. Zhu, Y.Q. Zhang,
\newblock{An efficient transport scheme for multimedia over wireless internet,}
\newblock{IEEE Int. Conf. on 3G wireless and beyond (3Gwireless), 2001}

\bibitem{RS}
K. Ramin, K. Sabmatian,
\newblock{Evaluation of packet error rate in wireless networks,}
\newblock{7th IEEE/ACM Int. Sym. on Mobility, Analysis and Simulation of Wireless Mobile Systems (MSWIM), 2004}

\bibitem{GAGM}
D. Gomez, R. Aguero, M. Garcia-Arranz, L. Munoz,
\newblock{Replication of the Bursty Behavior of Indoor WLAN Channels,}
\newblock{Proc. of the 6th Int. ICST Conf. on Simulation Tools and Techniques, (2013) pp. 219-226}








\end{thebibliography}


\end{document}